\documentclass[pre,aps ,superscriptaddress,showpacs]{revtex4-1}
\usepackage{natbib}
\usepackage{amsmath}
\usepackage{epsfig}
 \usepackage{color}
\begin{document}

\title{Characteristic Functions Based on Quantum Jump Trajectory  }
\author{Fei Liu}
\email[Email address: ]{feiliu@buaa.edu.cn}
\affiliation{School of Physics and Nuclear Energy Engineering, Beihang University, Beijing 100191, China}
\author{Jingyi Xi}

\affiliation{Academy for Advanced Interdisciplinary Studies, Peking University, Beijing, China}
\date{\today}

\begin{abstract}
{Characteristic functions (CFs)  provide a very efficient method for evaluating the probability density functions of stochastic thermodynamic quantities and investigating their statistical features in quantum master equations (QMEs). A conventional procedure for obtaining these functions is to resort to a first-principles approach; namely, the evolution equations of the CFs of the combined system and its environment are obtained and then projected into the degrees of freedom of the system. However, the QMEs can be unraveled by a quantum jump trajectory. Thermodynamic quantities such as the heat, work, and entropy production can be well defined along a trajectory. Hence, on the basis of the notion of a trajectory, can we straightforwardly derive these CFs, e.g., their evolution equations? This is essential to establish the self-contained stochastic thermodynamics of a QME. In this paper, we show that it is indeed plausible and also simple. Particularly, these equations are fully consistent with those obtained by the first-principles method. Our results have practical significance; they indicate that the quantum fluctuation relations could be verified by more realistic photocounting experiments. } 

\end{abstract}
\maketitle

\section{Introduction}
\label{section1}
In the past two decades, there has been growing interest~\cite{Mukamel2003,DeRoeck2004,DeRoeck2007,Esposito2009,Horowitz2013,
Esposito2006,Crooks2008,Talkner2009,Horowitz2012,Liu2012,Chetrite2012,Jaksic2013,Hekking2013,Horowitz2013,Leggio2013,
Liu2014,Liu2014a,Silaev2014,Suomela2014,Suomela2015,Manzano2015,Gasparinetti2014,Cuetara2015,Alonso2016,Gong2016} in the stochastic thermodynamics of nonequilibrium quantum processes that can be described by Lindblad quantum master equations (QMEs)~\cite{Davies1974,Lindblad1976,Gorini1976}. These studies include the definitions of stochastic thermodynamic quantities, their statistical features and computations, experimental measurements, etc.  This research was initially inspired by theoretical efforts that extend the classical fluctuation relations~\cite{Bochkov1977,Evans1993,Gallavotti1995,Jarzynski1997,Kurchan1998,
Lebowitz1999,Maes1999,Crooks2000,Hatano2001,Seifert2005,Kawai2007} into the quantum regime~\cite{Bochkov1977,Kurchan2000,Tasaki2000,
Yukawa2000,Allahverdyan2005,Talkner2007,Andrieux2008,Campisi2011,Batalhao2014,ShuomingAn2015,Jarzynski2015}. The practical possibility of manipulating and controlling quantum systems~\cite{Boixo2014,Kosloff2013,Gemmer2005} has further boosted the enthusiasm of researchers in this field. It is not occasional to devote considerable amount attention to QMEs. On the one hand, these equations have solid mathematical and physical foundations~\cite{Breuer2002,Alicki2010,Rivas2012}. On the other hand, in the statistical physics community, there has a long tradition of studying irreversible thermodynamics by using QMEs~\cite{Spohn1978,Davies1978,Alicki1979,Kosloff2013}. Although most of these previous results have been obtained from the point of view of an ensemble or average, they provide a very valuable reference for studies of the stochastic behaviors of these quantum systems.

A powerful concept for investigating stochastic thermodynamic quantities is the characteristic function (CF)~\cite{Gardiner1983,Risken1984}. A CF has been employed in the work equalities of closed quantum systems~\cite{Campisi2011} and was soon extended to the case of QMEs~\cite{Talkner2007,Esposito2009}~\footnote{In some studies, e.g., Refs.~\cite{Esposito2009},~\cite{Garrahan2010},~\cite{Gasparinetti2014},~\cite{Cuetara2015}, and~\cite{Pigeon2015}, the moment-generating function rather than the CF was used. If all of the moments of a stochastic quantity exist and are finite, which we assume throughout this article, there are no essential differences between two functions. A moment-generating function may be simply regarded as a CF evaluated on the imaginary axis~\cite{Gardiner1983}.}. Previous results have shown that this concept is very efficient for analyzing fluctuation properties and the computation of stochastic quantities~\cite{Kurchan1998,Lebowitz1999,Imparato2007,Talkner2009,Esposito2009,Silaev2014,Gasparinetti2014,Liu2014,Liu2014a,Cuetara2015} and indispensable in the large-deviation formalism~\cite{Touchette2008,Esposito2009,Garrahan2010,Zinidarifmmodeheckclseci2014,Pigeon2015}. In QMEs, a conventional route for obtaining CFs is a ``first-principles'' method~\cite{Talkner2009,Esposito2009,Silaev2014,Gasparinetti2014,Suomela2014,Cuetara2015}. Namely, one first regards the system of interest and its surrounding reservoirs as a composite system. By defining the CFs of the stochastic heat or work in this closed quantum system on the basis of the two-energy measurement (TEM) scheme~\cite{Kurchan2000}, one can then construct the time-evolution equations for these CFs~\cite{Esposito2009}. The last step is to project these equations into the degrees of freedom of the system. Since many key approximations such as the weak coupling limit and rotating wave approximation are involved, this step is very analogous to the derivations of QMEs~\cite{Breuer2002}. There is no doubt that this method of obtaining CFs is rigorous in mathematics and also reasonable in physics. Moreover, it even has some advantages, e.g., the exhibition of non-Markovian effects. However, from the point of view of establishing the self-contained stochastic thermodynamics of QMEs, this route is not very satisfactory. First, we are given a QME {\it beforehand}  and then study its entropy production or the energy exchanges between the system and the reservoirs. Logically, the introduction of a composite system should not be essential. This point is understood more clearly if one recalls the stochastic thermodynamics of classical stochastic processes~\cite{Seifert2011}. For example, consider the statistical features of the heat of a Brownian particle moving in a fluid. The equation of motion of the particle is usually described by the Fokker--Planck  equation whose role is very analogous to a QME. The energies of the system consisting of the particle and its surrounding fluid would rarely be considered. In addition, not all QMEs have microscopic foundations; some of them have been proposed phenomenologically~\cite{Rousochatzakis2005,Cai2010,Kosloff2013}. Under this circumstances, the first-principles method might lose microscopic basis.

An alternative strategy for studying the stochastic thermodynamics of QMEs is to apply the fact that these equations can be unraveled into a quantum jump trajectory (QJT)~\cite{Carmichael1993,Plenio1998,Breuer2002,Wiseman2010}. Along each QJT, the heat, work, and entropy production can be well defined~\cite{Breuer2003,DeRoeck2007,Derezinski2008,Crooks2008,Horowitz2012,Hekking2013,Leggio2013,Liu2014,Liu2014a,Suomela2015,Elouard2015}. Since the occurrence of a trajectory has a conventional probability interpretation, the CFs shall be naturally defined as well. In order to analyze the statistical features of these thermodynamic quantities, we still need to obtain the time-evolution equations of these functions. However, to the best of our knowledge, there are few papers that have carried out this critical step. Previously, for special time-independent QMEs, De Roeck et al.~\cite{DeRoeck2007,Derezinski2008} argued that the CF of the heat defined by a QJT is equivalent to the CF defined by the first-principles method. However, they did not derive any time-evolution equations. Very recently, for two specific time-dependent QMEs, one of the authors developed a method for calculating the work by solving the backward-time-evolution equations of the CF of the work~\cite{Liu2014,Liu2014a,Liu2016}. In addition to the less frequently used backward time, the introduction of the auxiliary time-reversed QMEs therein also restricted the region of application of this method. In this paper, we attempt to thoroughly overcome this issue. Rather than focusing on some special models, our discussions are focused on a QME whose form is sufficiently general to cover the various QMEs frequently found in the literature. We show that there are not any special difficulties for obtaining the time-evolution equation of these CFs. Importantly, these CFs are also consistent with those obtained by the previous first-principles method. 

The paper is organized as follows. In Sec.~\ref{section2} we unify a variety of QMEs into a general formula. We point out that this formula may be interpreted by the notion of the QJT. The essential notation is also provided. In Sec.~\ref{section3}, after an overview of the thermodynamic quantities at the trajectory level and the clarification of their applicable regions, we show that their CFs can be always evaluated by taking the trace over the operators. Particularly, their equations of motion can be straightforwardly derived and have more than one form. The statistical properties of these quantities are also discussed from the point of view of the CFs. In Sec.~\ref{section5}, we use a concrete Floquet QME to illustrate our results. Section~\ref{section6} presents the conclusions of this paper.

\section{QME and QJT}
\label{section2}
There exists a variety of QMEs in stochastic thermodynamics. We roughly divide them into three types~\footnote{The closed quantum systems may be regarded as a type of QME. Because of trivialness, we do not consider it. }. The first type is the standard master equations~\cite{Davies1974,Lindblad1975,Gorini1976,Rivas2012}. These time-independent equations have been mainly applied to issues related to how a system relaxes into a thermal equilibrium state~\cite{Breuer2002,Alicki2010} or nonequilibrium steady state~\cite{DeRoeck2004,Derezinski2008,Esposito2009,Garrahan2010,Zinidarifmmodeheckclseci2014,Schaller2014}. In contrast, the equations of the second type have time-dependent coherent dynamics, whereas the dissipative parts are static~\cite{Geva1994,Hekking2013,Liu2014,Silaev2014}. They have often been used in quantum optics~\cite{Breuer2002,Carmichael1993,Wiseman2010}, e.g., a two-level atom interacting with a radiation field and driven by a classical time-varying electric field~\cite{Mollow1975}. The physical validity of these equations is ensured if the externally driven field is so weak that its effect on its environment is negligible. The equations of the last type fully depend on the time. Typical examples include the adiabatically driven QMEs~\cite{Davies1978,Alicki1979,Albash2012,Horowitz2012,Liu2014a,Suomela2015} and periodic Floquet QME~\cite{Bluemel1991,Kohler1997,Breuer1997,Szczygielski2013,Gasparinetti2014,Cuetara2015}. Although the applicable regions of these three types of QMEs are very distinct~\cite{Alicki2006}, they can be formally unified as specific cases of the following QME:
\begin{eqnarray}
\label{masterequation}
\partial_t\rho&=&-\frac{i}{\hbar}[H(t),\rho]+\sum_{\omega_t}\gamma(\omega_t)\left[A(\omega_t,t)\rho A^\dag (\omega_t,t)-\frac{1}{2}\left\{ A^\dag(\omega_t,t)A(\omega_t,t),\rho\right \}\right]\nonumber\\
&=&{\cal L}(t)\rho,
\end{eqnarray}
where $H(t)$ is the Hamiltonian of the system, $\rho(t)$ is the reduced density matrix of the system, $A(\omega_t,t)$ are the Lindblad operators, and $A^\dag(\omega_t,t)=A(-\omega_t,t)$. In this paper, we further assume that there is only one heat reservoir surrounding the system at the equilibrium temperature $T$. Eq.~(\ref{masterequation}) is not the most general. We may add more dissipative terms to account for the complex interactions between the system and the reservoir or for the presence of multiple reservoirs, e.g., as in Ref.~\cite{Pigeon2015}. Nevertheless, it is sufficient to show our formulas and results. The Lindblad operators mean that they are the eigenoperators  of an operator, ${\cal H}(t)$, i.e.,
\begin{eqnarray}
\label{eigenopator}
&&[{\cal H}(t),A^\dag(\omega_t,t)]=\hbar\omega_t A^\dag(\omega_t,t),\nonumber\\
&&[{\cal H}(t),A(\omega_t,t)]=-\hbar\omega_t A(\omega_t,t).
\end{eqnarray}
The coefficients $\hbar\omega_t$ are the differences between the eigenvalues of ${\cal H}(t)$. They are positive or negative but always occur in pairs. We note that ${\cal H}(t)$ may or may not be a physical Hamiltonian, which depends on concrete models. For instance, in a system driven by a weak field, ${\cal H}(t)$ is the unperturbed Hamiltonian $H_0$ of the system, and $\hbar\omega_t$ are the differences between the energy eigenvalues of $H_0$~\cite{Breuer2002,Liu2014}. In an adiabatically driven system, ${\cal H}(t)$ is the system's Hamiltonian $H(t)$, while $\hbar\omega_t$ are the differences between the instantaneous energy eigenvalues of $H(t)$~\cite{Albash2012,Liu2014a}. In a periodic Floquet QME, ${\cal H}(t)$ is the Floquet Hamiltonian, $H_F=H(t)-i\hbar\partial_t$. In this case, the commutator in Eq.~(\ref{eigenopator}) must be understood in the Sambe space~\cite{Sambe1973}, while $\hbar\omega_t$ are the differences between the quasienergies of the Floquet Hamiltonian~\cite{Shirley1965,Zeldovich1967}.
The last component of the QME in Eq.~(\ref{masterequation}) is related to the rates $\gamma(\omega)$ ($>0$). A conventional assumption is that they satisfy the detailed balance condition
\begin{eqnarray}
\label{detailedbalance}
\gamma(\omega)=e^{\hbar \omega/k_BT}\gamma(-\omega).
\end{eqnarray}
This is essential for the validity of a variety of fluctuation relations~\cite{Mukamel2003,DeRoeck2004,DeRoeck2007,Esposito2009,Horowitz2013,
Esposito2006,Crooks2008,Talkner2009,Horowitz2012,Liu2012,Chetrite2012,Jaksic2013,Hekking2013,Horowitz2013,Leggio2013,
Liu2014,Liu2014a,Silaev2014,Suomela2014,Suomela2015,Gasparinetti2014,Cuetara2015,Alonso2016}. It is worthwhile to emphasize that either the seemingly complex descriptions of ${\cal H}(t)$ or the detailed balance condition have nothing to do with our formalism for the CFs.

The solution of Eq.~(\ref{masterequation}) is formally written as $\rho(t)=G(t,t_0)(\rho(t_0))$, where the superpropagator is
\begin{eqnarray}
\label{systemsuperpropagator}
G(t,t_0)={\cal T}_\leftarrow e^{ \int_{t_0}^t d\tau \cal{L}(\tau)},
\end{eqnarray}
where ${\cal T}_\leftarrow $ is the chronological time-ordering operator. On the other hand, $\rho(t)$ can be also interpreted as a statistical average of an ensemble of wave vectors~\cite{Carmichael1993,Plenio1998,Breuer2002,Wiseman2010}. To understand this point clearly, it is insightful to apply the Dyson series to Eq.~(\ref{masterequation}). This formalism was initially developed for time-independent QMEs~\cite{Carmichael1993,Wiseman1993,Kist1999}. Nevertheless, its extension to the current QME is almost trivial. First, we rewrite Eq.~(\ref{masterequation}) as
\begin{eqnarray}
\label{peturbedmastereq}
\partial_t \rho={\cal L}_{0}(t)\rho+\sum_{\omega_t}J(\omega_t,t)\rho,
\end{eqnarray}
where the superoperators ${\cal L}_0$ and $J_\omega$ are
\begin{eqnarray}
\label{continuouspart}
&&{\cal L}_{0}(t)\rho=-\frac{i}{\hbar}[H(t),\rho]-\frac{1}{2}\sum_{\omega_t}\gamma(\omega_t)\left\{A^\dag(\omega_t,t)A(\omega_t,t),\rho\right\},\\
\label{jumppart}
&&J({\omega_t},t)\rho=\gamma(\omega_t)A(\omega_t,t)\rho A^\dag(\omega_t,t),
\end{eqnarray}
respectively~\footnote{Note that the separation of the QME~(\ref{masterequation}) is not unique. Different separations may correspond different experimental monitoring schemes of the open system~\cite{Kist1999}. }. Applying the Dyson series to Eq.~(\ref{peturbedmastereq}), we obtain the following alternative formal solution of $\rho(t)$:
\begin{eqnarray}
\label{formalsolutionrho}
\rho(t)&=&G_0(t,t_0)\left[\rho(t_0)\right]\nonumber \\
&+& \sum_{N=1}^\infty\sum_{\{\omega_{i}\}} \left(\prod_{i=N}^1 \int_{t_0}^{t_{i+1}}\right)   \left(\prod_{i=N}^1dt_i\right) G_0(t,t_N)J(\omega_{t_N},t_N)G_0(t_N,t_{N-1})\cdots J(\omega_{t_1},t_1)G_0(t_1,t_0)\left[\rho(t_0)\right]\nonumber\\
&\doteq&\int_C {\cal D}(t)\hspace{0.1cm} 
G_0(t,t_N)J(\omega_{t_N},t_N)G_0(t_N,t_{N-1})\cdots J(\omega_{t_1},t_1)G_0(t_1,t_0)\left[\rho(t_0)\right],
\end{eqnarray}
where $t_{N+1}$$=$$t$, $\{\omega_{i}\}$$=$$\{\omega_{t_N},\cdots,\omega_{t_1}\}$, the summations are over all possible $\omega_{t_i}$ at time $t_i$, and the superpropagator is
\begin{eqnarray}
G_0(t,t')={\cal T}_{\leftarrow}e^{\int_{t'}^{t}d\tau{\cal L}_0(\tau)}.
\end{eqnarray}
The reader is reminded that these superoperators act on all terms on their right-hand side. For simplification of the notation, we used the abbreviation ${\cal D}(t)$ and the subscript {\it C} to denote these integrals and summations with respect to all possible arrangements. The structures of Eqs.~(\ref{continuouspart}) and~(\ref{jumppart}) show that, if the initial density matrix $\rho(t_0)$ is a pure state, the action of the integrand of Eq.~(\ref{formalsolutionrho}) always preserves this purity during the entire procedure, that is, a quantum trajectory in the Hilbert space of the system is generated~\cite{Carmichael1993}. Importantly, a further argument~\cite{Carmichael1993,Breuer2002,Wiseman2010} shows that the classical probability of observing such a trajectory in the time interval $(t_0,t)$ that has an initial wave vector of $|\psi_m\rangle$, undergoes $N$ jumps at increasing times $t_i$ ($i$$=$$1$, $\cdots$, $N$) with an order of jumps $\{\omega_i\}$, and finally arrives at the wave vector $|\phi_n\rangle$ is as follows:
\begin{eqnarray}
\label{trajprob}
dP_{n|m}\{\omega_{i}\}=\left(\prod_{i=N}^1dt_i\right){\rm Tr}\left[|\phi_n\rangle \langle \phi_n| G_0(t,t_N)J(\omega_{t_N},t_N)G_0(t_N,t_{N-1})\cdots J(\omega_{t_1},t_1)G_0(t_1,t_0)(|\psi_m \rangle \langle \psi_m|)\right].
\end{eqnarray}
If no jump occurs, the probability is simply
\begin{eqnarray}
\label{trajprob0}
dP_{n|m}\{\omega_{i}\}= {\rm Tr}\left[|\phi_n\rangle \langle \phi_n|G_0(t,t_0)(|\psi_m \rangle \langle \psi_m|)\right].
\end{eqnarray}
Unless otherwise stated, $|\psi_m\rangle$ and $|\phi_n\rangle$ are assumed to be the eigenvectors of some physical quantities with the quantum numbers $m$ and $n$, respectively.

\section{CFs and time-evolution equations }
\label{section3}
\subsection{Thermodynamic quantities}
The physical realization of a QJT is that an open quantum system is continuously monitored by an external photon detector~\cite{Carmichael1993,Wiseman1993,Kist1999}. This mechanism interprets an action of $J(\omega_t,t)$ on a pure state $|\varphi(t)\rangle\langle\varphi(t)|$ at the time $t$ as a jump of the wave vector $|\varphi(t)\rangle$ to a new wave vector $A(\omega_t,t)|\varphi(t)\rangle/ \|A(\omega_t,t)|\varphi(t)\rangle\|$. Importantly, this jump also accompanies an energy exchange $|{\hbar\omega_t}|$ between the system and the reservoir, which is recorded by the detector. If
the sign of ${\hbar \omega_t}$ is positive, the energy is released into the reservoir; otherwise, it is absorbed from the reservoir~\cite{Mollow1975,Carmichael1993,Breuer1997,Breuer2002}. Hence, along a trajectory starting from the vector $|\psi_m\rangle$ and with $N$ jumps, $\{\omega_i\}$, we define the heat produced by the system as~\cite{Breuer2003,Derezinski2008,Garrahan2010,Horowitz2012,Leggio2013,Hekking2013,Liu2014,Liu2014a,Suomela2015,Gong2016}
\begin{eqnarray}
\label{heatdef}
Q_{n|m}\{\omega_i\}=\sum_{i=1}^N \hbar\omega_{t_i}.
\end{eqnarray}
This definition does not depend on the concrete initial and final wave vectors. We only require that the initial density matrix is a mixture of pure states, i.e., $\rho(t_0)=\sum p_m|\psi_m\rangle\langle \psi_m|$, where $p_m$ is the probability of finding the state $|\psi_m\rangle$. These wave vectors could or could not be orthogonal each other.

To define the stochastic work done by some external devices, a TEM~\cite{Esposito2009,Campisi2011} of the system must be performed. Assume that the system's Hamiltonian $H(t)$ has instantaneous eigenvectors $|\varepsilon_n(t)\rangle$ with discrete eigenvalues $\varepsilon_n(t)$, $n=1,\cdots$. By performing TEMs of the system at the beginning and ending of the quantum process, we define the work along a QJT with $N$ jumps, $\{\omega_i\}$, and starting from the state $|\varepsilon_m(t_0)\rangle$ and ending at the state $|\varepsilon_n(t)\rangle$ as~\cite{Horowitz2012,Hekking2013,Liu2014a,Gong2016}
\begin{eqnarray}
\label{workdef}
W_{n|m}\{ \omega_i \}=\varepsilon_n(t)-\varepsilon_m(t_0)+
\sum_{i=1}^N \hbar\omega_{t_i}.
\end{eqnarray}
Eq.~(\ref{workdef}) is in fact the first law of thermodynamics under the notion of a QJT. However, because of the energy measurement at time $t_0$, the original density matrix $\rho(t_0)$ is projected into a new one given by
\begin{eqnarray}
\label{projecteddensitymatrix}
\rho'(t_0)&=&\sum_m  \langle \varepsilon_m(t_0)|\rho(t_0)|\varepsilon_m(t_0)\rangle |\varepsilon_m(t_0)\rangle\langle \varepsilon_m(t_0)|\nonumber\\
&=&\sum_m  \rho_{mm}(t_0) |\varepsilon_m(t_0)\rangle\langle \varepsilon_m(t_0)|,
\end{eqnarray}
where $\rho_{mm}(t_0)$ is the probability of finding the wave vector $|\varepsilon_m(t_0)\rangle$~\cite{Esposito2009,Campisi2011}. In addition to Eq.~(\ref{workdef}), stochastic work may also be  defined for one part of the system~\cite{Liu2014}. For instance, in the QMEs of the second type, their Hamiltonian usually has two terms, $H(t)=H_0+H_I(t)$, where $H_0$ is the time-independent bare Hamiltonian of the atoms and $H_I(t)$ is the interaction between the atoms and external classical fields. The TEM may be performed on $H_0$ rather than the entire $H(t)$. Then an alternative work is defined as Eq.~(\ref{workdef}); however, the eigenvectors and eigenvalues therein are replaced by those of the bare Hamiltonian $H_0$~\cite{Liu2012,Liu2014}. Eq.~(\ref{projecteddensitymatrix}) is modified accordingly. Although these two types of work are distinct in physics~\cite{Campisi2011a,Kosloff2013,Liu2014b}, their formulas are very similar. Hence, we use a notation $H'(t)$ to represent their different Hamiltonian, and $|\varepsilon_m(t)\rangle |$ and $\varepsilon_n(t)$ in Eqs.~(\ref{workdef}) and~(\ref{projecteddensitymatrix}) are understood as the eigenvectors and eigenvalues of $H'(t)$, respectively. Note that these various definitions of work in the same system are also present in the classical situation~\cite{Jarzynski2007}.

The last thermodynamic quantity is the total entropy production. Considering that the density matrix $\rho(t)$ is Hermitian, it must have a diagonal form with respect to an {orthonormal} basis, i.e., $\rho(t)$$=$$\sum_n\lambda_n(t)|\lambda_n(t)\rangle\langle \lambda_n(t)|$~\cite{Sakurai1994}. The probability $\lambda_n(t)$ of finding the state $|\lambda_n(t)\rangle$ is also the eigenvalue of $\rho(t)$ itself. If we assume that a QJT starts from the wave vector $|\lambda_m(t_0)\rangle$, jumps $N$ times with $\{\omega_i\}$, and ends at the wave vector $|\lambda_n(t)\rangle$, we can define the total entropy production along this trajectory as
\begin{eqnarray}
\label{entropyproddef}
S_{n|m}\{\omega_i\}=k_B[-\ln \lambda_n(t)+\ln \lambda_m(t_0)]+\frac{1}{T}\sum_{i=1}^N \hbar\omega_{t_i}.
\end{eqnarray}
This is a simple quantum extension of the classical trajectory entropy~\cite{Seifert2005}. We must emphasize that this definition is distinct from the classical one; here, $-k_B\ln\lambda_m(t)$ is related to the von Neumann entropy instead of the Shannon entropy that was used in the classical case~\cite{Breuer2002}.

\subsection{Heat}
On the basis of the above notation, we now write the CF for the heat as follows:
\begin{eqnarray}
\label{CFheatdef}
\Phi(\xi)&=&\left\langle \exp(i\xi Q)\right\rangle\nonumber\\
&=&\sum_{n,m}\int_CdP_{n|m}\{\omega_i\}p_m\exp\left[i\xi Q_{n|m}\{\omega_i\}\right].
\end{eqnarray}
The first equation is shorthand for the average over QJTs. Substituting the probability formulas in Eqs.~(\ref{trajprob}) and~(\ref{trajprob0}) and the heat definition in Eq.~(\ref{heatdef}) into Eq.~(\ref{CFheatdef}) and rearranging, we obtain
\begin{eqnarray}
\label{CFheatmiddle}
\Phi(\xi)={\rm Tr}\left[\int_C {\cal D}(t) \hspace{0.1cm}
 G_0(t,t_N)e^{i\xi\hbar\omega_{t_N}}J(\omega_{t_N},t_N)G_0(t_N,t_{N-1})\cdots e^{i\xi\hbar\omega_{t_1}}J(\omega_{t_1},t_1)G_0(t_1,t_0)\left[\rho(t_0)\right]\right].
\end{eqnarray}
We immediately find that the entire term in the square brackets is almost the same as the formal solution of $\rho(t)$; see Eq.~(\ref{formalsolutionrho}). The only difference is that each jump superoperator $J(\omega_{t},t)$ is multiplied by the ``phase'' factor $\exp(i\xi\hbar\omega_t)$. Hence, without further derivation, this analogy results in the following alternative expression of the CF:
\begin{eqnarray}
\label{CFheatDensitymatrixForm}
\Phi(\xi)={\rm Tr}[\hat{{\rho}}(t,t_0;\xi)]={\rm Tr}\left[\check{G}(t,t_0;\xi)(\rho(t_0))\right],
\end{eqnarray}
where the new operator $\hat{{\rho}}(t,t_0;\xi)$ satisfies the time-evolution equation given by
\begin{eqnarray}
\label{heatmasterequation}
\partial_t \hat{{\rho}}&=&-\frac{i}{\hbar}[H(t), \hat{{\rho}}]+\sum_{\omega_t}\gamma (\omega_t)\left[e^{i\xi\hbar\omega_t}A(\omega_t,t) \hat{{\rho}} A^\dag(\omega_t,t)-\frac{1}{2}\left\{ A^\dag(\omega_t,t)A(\omega_t,t), \hat{{\rho}}\right \}\right]\nonumber\\
&=&\check{\cal{L}}(t;\xi)\hat{\rho},
\end{eqnarray}
and its initial condition is $\rho(t_0)$. This is the central result of this paper. Considering that Eq.~(\ref{heatmasterequation}) will be repeatedly used but with different initial conditions below, we specifically define its superpropagator
\begin{eqnarray}
\check{G}(t,t_0;\xi)={\cal T}_\leftarrow e^{{\int_{t_0}^t d\tau\check{\cal{L}}(\tau;\xi)}}.
\end{eqnarray}

We have several comments regarding the time-evolution equation. Firstly, in a concrete QME with a special Hamiltonian and Lindblad operators,  Eq.~(\ref{heatmasterequation}) reduces to the previous results~\cite{Esposito2009,Garrahan2010,Silaev2014,Gasparinetti2014,Zinidarifmmodeheckclseci2014,Cuetara2015}. All of them were obtained by the first-principles method mentioned in Sec.~\ref{section1}. Obviously, this is dramatically different from our method that is completely based on a QJT. In addition, it is worthwhile to point out that the earliest version of Eq.~(\ref{heatmasterequation}) is credited to Mollow~\cite{Mollow1975}. He investigated the probability distribution of the number of photons of a two-level atom that is driven by a weak classical field and simultaneously interacts with a bath of modes of a radiation field. At that time, the notion of a QJT was still in its infancy, and emission or absorption of photons was not interpreted as ``heat.'' Hence, it is not very surprising to see that the preceding derivations carried out by Mollow are far more complex. Secondly, the CF in Eq.~(\ref{CFheatDensitymatrixForm}) provides a convenient way to obtain the operator expressions of the moments of the heat. To clarify this point, we rewrite Eq.~(\ref{heatmasterequation}) as
\begin{eqnarray}
\label{peturbedmastereqheat}
\partial_t \hat{{\rho}}&=&{\cal{L}}(t)\hat{\rho}+\sum_{\omega_t}\left(e^{i\xi\hbar\omega_t}-1\right)J(\omega_t,t) \hat{{\rho}}\nonumber\\
&=&{\cal{L}}(t)\hat{\rho}+{\cal Q}_{\xi}(t)\hat{{\rho}}.
\end{eqnarray}
This new form is very analogous to Eq.~(\ref{peturbedmastereq}). We may apply the Dyson's series again and obtain its formal solution
\begin{eqnarray}
\hat{{\rho}}(t,t_0;\xi)=\int_C {\cal D}(t)\hspace{0.1cm} G(t,t_N){\cal Q}_{\xi}(t_N)G(t_N,t_{N-1})\cdots {\cal Q}_{\xi}(t_1)G(t_1,t_0)[\rho(t_0)],
\end{eqnarray}
where $G(t_2,t_1)$ is the superpropagator in Eq.~(\ref{systemsuperpropagator}). Substituting the solution into the CF of the heat and performing a Taylor expansion in terms of $\xi$, we easily obtain the moments of the heat, e.g., the first two moments,
\begin{eqnarray}
\label{firstsecondmoentheat}
&&\langle Q \rangle=\int_{t_0}^{t}dt_1 \sum_{\omega_{t_1}} \hbar\omega_{t_1} \gamma(\omega_{t_1} ) \left\langle A^
\dag (\omega_{t_1},t_1)A(\omega_{t_1},t_1)\right\rangle,\\
&&\langle Q^2 \rangle=\int_{t_0}^{t} dt_1\sum_{\omega_{t_1}}(\hbar\omega_{t_1})^2 \gamma(\omega_t)\left\langle A(\omega_{t_1},t_1)A^\dag (\omega_{t_1},t_1)\right\rangle \nonumber\\
&&\hspace{1.cm}+2\int_{t_0}^{t}dt_1\int_{t_0}^{t_1} dt_2\sum_{\omega_{t_2},\omega_{t_1}} \hbar\omega_{t_2}\hbar\omega_{t_1}\gamma(\omega_{t_2})\gamma(\omega_{t_1})\left\langle A(\omega_{t_1},t_1)A(\omega_{t_2},t_2)A^\dag (\omega_{t_2},t_2)A^\dag (\omega_{t_1},t_1)\right\rangle.
\end{eqnarray}
Here, we have used the definition of the multitime correlation function of operators~\cite{Gardiner2004}; see Appendix A. Lastly, although the presence of Eq.~(\ref{heatmasterequation}) does not matter to Eq.~(\ref{detailedbalance}), this detailed balance condition is indeed critical to ensure the fluctuation relations. Under this condition, the superoperator $\check{G}$ possesses the following important property: \begin{eqnarray}
\label{propertyofsuperpropagator}
\check{G}(t,t_0;i\beta)(I)=0,
\end{eqnarray}
where $\beta$ is the inverse temperature $1/k_BT$, and $I$ is the identity operator. The proof is straightforward. Considering that the presence of positive and negative $\omega_t$ is always in pairs and the Lindblads operator $A(-\omega_t,t)$ is equal to $A(\omega_t,t)^\dag$, we have $\check{\cal{L}}(t;i\beta)=0$. Eq.~(\ref{propertyofsuperpropagator}) is very useful for exploring intriguing fluctuation relations. For instance, we can obtain the following integral fluctuation relation for the heat:
\begin{eqnarray}
\label{FRheat}
\left\langle e^{-\beta Q}\right\rangle_{r}=1,
\end{eqnarray}
where the subscript $r$ indicates that the initial density matrix of the system is a completely random ensemble, e.g., in a $N$-level system, $\rho(t_0)$$=$$I/N$. The same equality has been found in a specific two-level system (TLS) of a second-type QME~\cite{Liu2016}. The current one is more general. The other two applications will be presented shortly.

\subsection{Work and total entropy production}
The CF of the work is
\begin{eqnarray}
\label{CFworkdef}
\Psi(\eta)&=&\left\langle \exp(i\eta W)\right\rangle'\nonumber\\
&=&\sum_{n,m}\int_C dP_{n|m}\{\omega_i\} \rho_{mm}(t_0)\exp\left[i\eta W_{n|m}\{\omega_i\}\right].
\end{eqnarray}
Here, we use a prime to indicate that the initial density matrix is $\rho'(t_0)$ instead of $\rho(t_0)$ itself. Substituting the probability formulas in Eqs.~(\ref{trajprob}) and~(\ref{trajprob0}) and the work definition in Eq.~(\ref{workdef}) into Eq.~(\ref{CFworkdef}), we have
\begin{eqnarray}
\label{CFworkmiddle}
\Psi(\eta)&=&{\rm Tr}\left[e^{i\eta H'(t)}\int_C {\cal D}(t) 
G_0(t,t_N)e^{i\eta\hbar\omega_{t_N}}J(\omega_{t_N},t_N)G_0(t_N,t_{N-1})\right.\nonumber\\
&&\hspace{5cm}\left.\cdots e^{i\eta\hbar\omega_{t_1}}J(\omega_{t_1},t_1)G_0(t_1,t_0)\left[e^{-i\eta H'(t_0)}\rho'(t_0)\right]\right].
\end{eqnarray}
In comparison with Eq.~(\ref{CFheatmiddle}), in addition to the fact that the initial density matrix is replaced by $\exp[-i\eta H(t_0)]\rho'(t_0)$, the other change is the presence of an additional operator $\exp[i\eta H'(t)]$. Hence, the CF of the work can be calculated in very similar manner to that the heat, i.e.,
\begin{eqnarray}
\label{CFworkfinal}
\Psi(\eta)={\rm Tr}\left[e^{i\eta H'(t)}\check{G}(t,t_0;\eta)\left(e^{-i\eta H'(t_0)}\rho'(t_0)\right)\right].
\end{eqnarray}
This result has been obtained for the second-type QME~\cite{Silaev2014} and Floquet QME~\cite{Cuetara2015} by applying the first-principles method, in which $H'(t)$ is equal to the bare Hamiltonian $H_0$ of the atoms and the entire Hamiltonian $H(t)$, respectively. Intriguingly, the detailed balance condition in Eq.~(\ref{detailedbalance}) and its consequence in Eq.~(\ref{propertyofsuperpropagator}) imply an important equality. If the projected density matrix $\rho'(t_0)$ happens to be the canonical distribution, i.e., $\exp[-\beta H'(t_0)]/Z(t_0)$, where $Z(t_0)$ is the instantaneous partition function at time $t_0$, i.e., ${\rm Tr}[\exp(-\beta H'(t_0))]$, the following equality appears:
\begin{eqnarray}
\label{workequality}
\left\langle e^{-\beta W}\right\rangle'=\frac{Z(t)}{Z(t_0)}.
\end{eqnarray}
This is nothing but the celebrated work equality for QMEs including Bochkov-Kuzovlev equality~\cite{Bochkov1977,Liu2014} and Jarzynski equality~\cite{Horowitz2012,Hekking2013,Liu2014a}.

The last CF is the total entropy production in Eq.~(\ref{entropyproddef}). Carrying out the same procedure as that for the heat or work, we obtain this CF as
\begin{eqnarray}
\label{CFentropyprod}
\Omega(\zeta)&=&\langle \exp(i\zeta S)\rangle\nonumber\\
&=&\sum_{n,m}\int_C dP_{n|m}\{\omega_i\} \lambda_m(t_0)\exp\left[i\zeta S_{n|m}\{\omega_i\}\right]\nonumber\\
&=&{\rm Tr}\left[\rho(t)^{-i\zeta k_B}\check{G}\left(t,t_0;\zeta/T\right)\left(\rho(t_0)^{i\zeta k_B}\rho(t_0)\right)\right].
\end{eqnarray}
Assuming the validity of the detailed balance condition in Eq.~(\ref{detailedbalance}) and choosing $\zeta=i/k_B$, we obtain the following fluctuation relation for the total entropy production:
\begin{eqnarray}
\label{entropyprodequality}
\left\langle e^{- S/k_B}\right\rangle=1.
\end{eqnarray}
According to Jensen's inequality, this equality implies that $\langle S\rangle\ge 0$, or the second law of thermodynamics for QMEs~\cite{Spohn1978a}. In contrast to the previous two fluctuation relations, Eqs.~(\ref{FRheat}) and~(\ref{workequality}), the current one is always true for an arbitrary initial condition.

\subsection{Alternative expressions of CFs}
\label{section4}
The above results give the impression that Eq.~(\ref{heatmasterequation}) holds a very special position. Either calculations or analyses of these CFs have to resort to this time-evolution equation. However, in the specific case of ${\cal H}(t)=H'(t)$, including the second-type QME~\cite{Liu2014} and adiabatically driven systems in third-type QMEs~\cite{Liu2014a}, we can construct an alternative time-evolution equation for the work that plays an analogous role as Eq.~(\ref{heatmasterequation}) for the heat. To clarify this point, we first regard the entire term in Eq.~(\ref{CFworkfinal}) as the operator $K(t,t_0;\eta)$. Differentiating $K(t,t_0;\eta)$ with respect to $t$ and applying Eq.~(\ref{heatmasterequation}) and the property of Lindblad operators in Eq.~(\ref{eigenopator}), we obtain the following closed evolution equation:
\begin{eqnarray}
\label{workmasterequationforward}
\partial_t K&=&{\cal L}(t)K+\partial_t \left(e^{i\eta H'(t)}\right)e^{-i\eta H'(t)}K-\frac{i}{\hbar}\left[e^{i\eta H'(t)},H(t)\right ]e^{-i\eta H'(t)}K \nonumber\\
&=&\hat{{\cal L}}(t;\eta)K,
\end{eqnarray}
and its initial condition is $\rho'(t_0)$. We particularly emphasize that the presence of Eq.~(\ref{workmasterequationforward}) is irrelevant to the detailed balance condition in Eq.~(\ref{detailedbalance}). This is another key result of this paper. If we re-solve the equation, the CF of the work is then alternatively calculated by
\begin{eqnarray}
\label{CFworkalternativeform}
\Psi(\eta)={\rm Tr}[K(t,t_0;\eta)]={\rm Tr}[\hat G(t,t_0;\eta)(\rho'(t_0))].
\end{eqnarray}
Here, the superpropagator is
\begin{eqnarray}
\hat{G}(t,t_0;\eta)={\cal T}_\leftarrow e^{  {\int_{t_0}^t d\tau\hat{\cal{L}}(\tau;\eta)}}.
\end{eqnarray}
We have several comments regarding Eq.~(\ref{workmasterequationforward}). First of all, this equation with Eq.~(\ref{CFworkalternativeform}) is just the quantum Feynman-Kac formula in the QMEs; see Appendix B. Secondly, analogous to the use of Eq.~(\ref{heatmasterequation}) for calculating the CFs of the work and total entropy production, we can use this equation with different initial conditions to calculate the CFs of the heat and total entropy production on the basis of the following formulas:
\begin{eqnarray}
&&\Phi(\xi)={\rm Tr}\left[e^{-i\xi H'(t)}\hat G(t,t_0;\xi)\left(e^{i\xi H'(t_0)}\rho(t_0)\right)\right],\\
&&\Omega(\zeta)={\rm Tr}\left[\rho(t)^{-i\zeta k_B}e^{-i\zeta H'(t)/T}\hat{G}\left(t,t_0;\zeta/T\right)\left(e^{i\zeta H'(t_0)/T}\rho(t_0)^{i\zeta k_B}\rho(t_0)\right)\right].
\end{eqnarray}
From a computation viewpoint for the condition of ${\cal H}(t)=H'(t)$, there are no significant differences between these two time-evolution equations. Hence, choosing one of them is only a question of personal taste.
Thirdly, if $H'(t)$ is the system's Hamiltonian $H(t)$, as in adiabatically driven QMEs~\cite{Liu2014a}, under the assumption of the detailed balance condition in Eq.~(\ref{detailedbalance}) and given $\eta=i\beta$, we note that Eq.~(\ref{workmasterequationforward}) is just the modified dynamics of the accompanying density matrix proposed by Chetrite and Mallick~\cite{Chetrite2012}. Its solution is then trivial and is as follows:
\begin{eqnarray}
K(t,t_0;i\beta)=\frac{\exp[-\beta H(t)]}{Z(t_0)}.
\end{eqnarray}
It was argued that these dynamics lead to a quantum Jarzynski equality. Nevertheless, they obtained it by simply projecting an analogous equation in the classical Langevin dynamics into the QME without any probability interpretation. This key ingredient was missing until explicitly establishing Eq.~(\ref{CFworkalternativeform}). Finally, Eq.~(\ref{workmasterequationforward}) has a backward time counterpart; see Appendix C. In fact, this version was proposed by one of the authors~\cite{Liu2014,Liu2014a}. However, to arrive at this result, he used auxiliary time-reversed QMEs and especially required the detailed balance condition, which makes the explanation of the backward equation obscure. Still, owing to Eq.~(\ref{CFworkalternativeform}), this ambiguity is thoroughly eliminated now.

\section{An example: Floquet QME of a two-level system}
\label{section5}
In this section, we use a concrete Floquet QME to demonstrate several previous results. This is a TLS with the Hamiltonian
\begin{eqnarray}
\label{FloquetHamiltonian}
H(t)=\frac{\hbar\omega_0}{2} \sigma_z +\frac{\hbar\Omega}{2}\left(\sigma_+ e^{-i\omega_L t}+\sigma_- e^{i\omega_L t}\right),
\end{eqnarray}
where $\omega_0$ is the frequency of the transitions between these two levels, $\Omega$ is the Rabi frequency, and $\omega_L$ is the frequency of the periodic external field. The thermodynamics of this typical QME has been intensively studied in the recent literature~\cite{Breuer2000,Kohn2001,Szczygielski2013,Langemeyer2014,Gasparinetti2014,Cuetara2015}. The Floquet basis vectors of this system are
\begin{eqnarray}
\label{Floquetbasis}
|u_{\pm}(t)\rangle =\frac{1}{\sqrt{2\Omega'}}
\left(\begin{array}{c}
 \pm \sqrt{\Omega'\pm\delta}\\
 e^{i\omega_L t}\sqrt{\Omega'\mp\delta},
\end{array}\right),
\end{eqnarray}
where $\Omega'=\sqrt{\delta ^2+\Omega^2}$, and the detuning parameter $\delta=\omega_0-\omega_L$. The corresponding quasienergies of these two vectors are
\begin{eqnarray}
\varepsilon_\pm=\frac{\hbar}{2}(\omega_L\pm \Omega'),
\end{eqnarray}
respectively. Assuming that $\omega_L-\Omega'>0$ and the coupling between the TLS and the heat reservoir is transverse~\cite{Breuer1997,Szczygielski2013,Langemeyer2014}, we may obtain three of the six Lindblad operators of $\omega=\omega_L$, $\omega_L-\Omega'$, and $\omega_L+\Omega'$ as
\begin{eqnarray}
\label{Lindbladoperators}
&&A(\omega_L,t)=\frac{\Omega}{2\Omega'}\left(|u_+(t)\rangle\langle u_+(t)|-|u_-(t)\rangle\langle u_-(t)| \right)e^{-i\omega_L t},\nonumber\\
&&A(\omega_L-\Omega',t)=\left(\frac{\delta-\Omega'}{2\Omega'} \right)|u_+(t)\rangle\langle u_-(t)|e^{-i\omega_L t},\\
&&A(\omega_L+\Omega',t)=\left(\frac{\delta+\Omega'}{2\Omega'} \right) |u_-(t)\rangle\langle u_+(t)| e^{-i\omega_L t},\nonumber
\end{eqnarray}
respectively. Note that these $\omega$ are indeed positive. Since they are also time-independent, we do not need the subscript $t$. The other three Lindblad operators $A(\omega,t)$ with $\omega=-\omega_L$, $-(\omega_L-\Omega')$, and $-(\omega_L+\Omega')$ are the adjoint operators of Eq.~(\ref{Lindbladoperators}).

Let us first check whether the first moment of the heat, $\langle Q\rangle$, in Eq.~(\ref{firstsecondmoentheat}) agrees with previous formulas~\cite{Langemeyer2014}. Substituting these Lindblad operators into Eq.~(\ref{firstsecondmoentheat}), we obtain its integrant as
\begin{eqnarray}
&&\hbar\omega_L \left(\frac{\Omega}{2\Omega'}\right)^2\left[\gamma(\omega_L)-\gamma(-\omega_L)\right]\nonumber\\
&&+\hbar(\omega_L-\Omega')\left(\frac{\delta-\Omega'}{2\Omega'}\right)^2
\left[\gamma(\omega_L-\Omega')p_--\gamma(-(\omega_L-\Omega'))p_+\right]\nonumber\\
&&+\hbar(\omega_L+\Omega')\left(\frac{\delta+\Omega'}{2\Omega'}\right)^ 2
\left[\gamma(\omega_L+\Omega')p_+-\gamma(-(\omega_L+\Omega'))p_-\right],
\end{eqnarray}
where $p_{\pm}=\langle u_{\pm}(t)|\rho(t)|u_{\pm}(t) \rangle $ are the diagonal elements of the reduced density matrix $\rho(t)$ in the Floquet basis, which have a probability meaning. We find that it is fully the same as the previous formula, e.g., Eq. (89) in Ref.~\cite{Langemeyer2014}. Second, since the time-evolution equation in Eq.~(\ref{heatmasterequation}) is derived on the basis of the notion of a QJT, it shall be interesting to verify this result by comparing the heat distributions resolved by the CF method and by directly simulating a QJT. To do this, we need expressions for the rates $\gamma(\omega)$. Assuming that the reservoir consists of an electromagnetic field in thermal equilibrium at a certain temperature $T$ and that the coupling between the TLS and the reservoir is a dipole interaction, these rates have the standard form~\cite{Breuer2002,Szczygielski2013}. If $\omega<0$,
\begin{eqnarray}
\gamma(\omega)=A|{\omega}|^3\frac{1}{e^{\hbar |\omega|/k_BT}-1};
\end{eqnarray}
otherwise, $\omega>0$, $\gamma(\omega)=e^{\hbar \omega/k_BT}\gamma(-\omega)$, where the coefficient $A$ depends on the dipole strength. Figure~(1) shows the numerical results. All computational details are given in Appendix D. We see that the agreement between these two methods is very impressive.
\begin{figure}\label{figure2}
\includegraphics[width=1.\columnwidth]{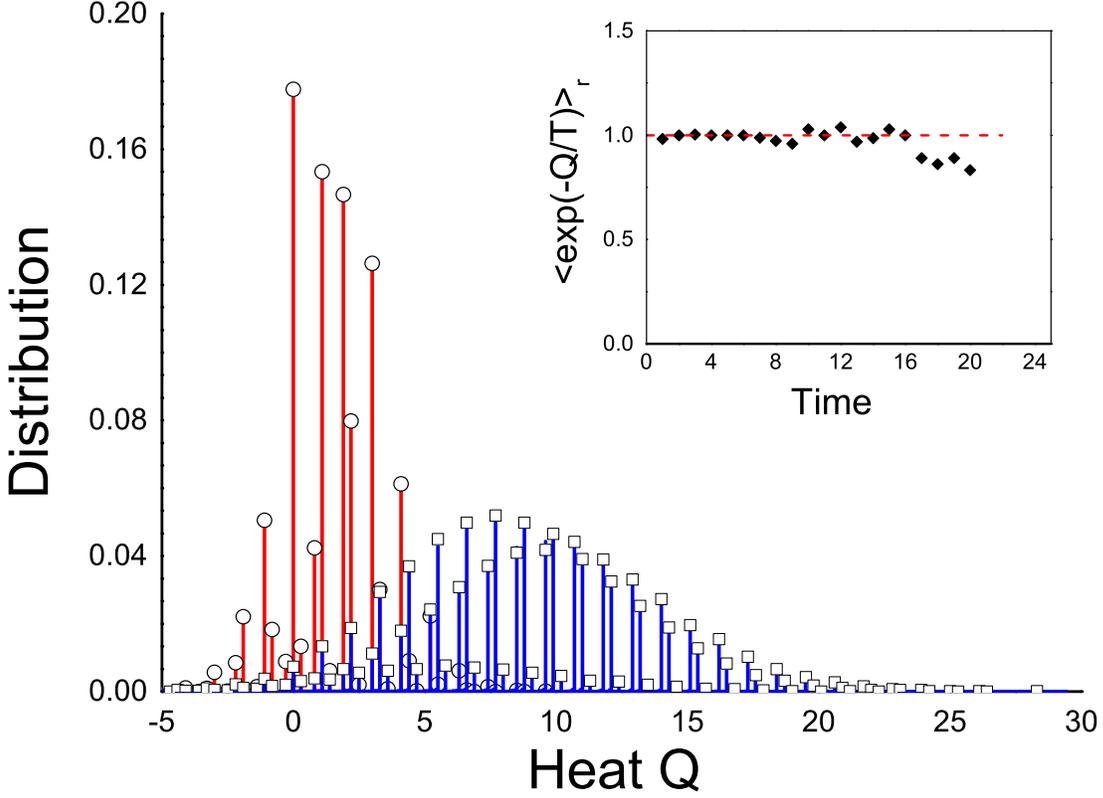}
\caption{The distributions of the heat (in units of $\hbar\omega_0$) of the Floquet QME of a TLS at $t=2\omega_0^{-1}$ (red lines and open circles) and $20\omega_0^{-1}$ (blue lines and open squares). The open symbols are calculated by simulating QJTs, whereas the vertical lines are obtained by solving Eq.~(\ref{heatmasterequation}) and performing an inverse Fourier transform. The initial density matrix is set to be the identity operator. The parameters are $\omega_L=1.1\omega_0$, $\Omega=0.8\omega_0$, $A=1.0\omega_0^{-2}$, and $T=1.0 \hbar\omega_0/k_B$. For convenience, we have let $k_B=1$, $\hbar=1$, and $\omega_0=1$. The inset shows the results of the left-hand side of Eq.~(\ref{FRheat}) calculated at different values of $t$ by simulating QJTs. The dashed line therein is a guide for the eyes.}
\end{figure}
The physical implication of these heat distributions have been discussed in detailed in Ref.~\cite{Gasparinetti2014}. Finally, we check the fluctuation relation~(\ref{FRheat}) of the heat using the QJT simulation data; see the inset of the Figure~(1). We see that this equality is satisfied if the duration of the process is relatively short, although there are apparent deviations as time increases. It is expected that the number of trajectories with negative heat production decreases dramatically as the duration of the process becomes longer; see the heat distribution at a larger time of $t=20$. We explain the reason for not using the data obtained by the CF method. Numerically solving Eq.~(\ref{heatmasterequation}) and performing the inverse Fourier transform always result in numerical errors. We unavoidably get some very small but nonzero probabilities at very negative heat for any duration. Since the exponential function is involved in Eq.~(\ref{FRheat}), these ``non-physical" probabilities can easily make the equality~(\ref{FRheat}) invalid, though they do not change the profile of the distributions. Compared with the sampling error, the numerical error is negligible in the QJT simulation.

\section{Discussion and conclusion}
\label{section6}
In this paper, we comprehensively investigated how to obtain the CFs of stochastic thermodynamic quantities in the QMEs by straightforwardly applying the notion of QJT. Our results show that their time-evolution equations can be obtained explicitly and in a simple manner. For the QMEs that have microscopic derivations, these time-evolution equations are fully consistent with those derived by the first-principles approach. Nevertheless, we need to point out that, our theory does not just provide an alternative derivation method. On the one hand, the above discussions obviously remind us that, for any QME that can be described by the general Eq.~(\ref{masterequation}), we may always establish its stochastic thermodynamics in a self-contained manner. Hence, our theory is valid even for the effective or phenomenological QMEs. On the other hand,
our theory definitely demonstrates that, quantum optics experiments, especially the photon-counting technique, could play very significant roles in studying the stochastic thermodynamics of quantum open systems. This has practical significance. Previous theories might give one an impression that, in order to carry out these studies, the TEM schemes would have to be performed on the combined system and reservoir. Now the notion of QJT opens a new avenue. Of course, we do not mean that, observing a QJT would be earlier than realizing the TEM schemes. However, the recent remarkable progress in experimentally measuring quantum trajectories has indeed set a high expectation~\cite{Murch2013,Sun2013,Vool2014,Campagne-Ibarcq2016}.

We conclude this paper by pointing out several extensions of the current theory. The first is to study scenarios containing multiple reservoirs or particle transport. So far, we have only been concerned with one reservoir with a thermal temperature, and particle exchanges are not allowed. Some literatures already considered this issue~\cite{Esposito2009,Szczygielski2013,Zinidarifmmodeheckclseci2014,Pigeon2015,Cuetara2015}. We see that the QMEs in these studies are analogous to Eq.~(\ref{masterequation}); however, more dissipative terms are present. Hence, the theory and computing method developed in this paper are useful in these situations. The second possible extension is to account for non-Markovian effects~\cite{Breuer2004,Hush2015}. Non-Markovian properties usually lead to a failure of the QME~(\ref{masterequation}) on which we heavily rely, e.g., the rates $\gamma(\omega)$ becomes negative~\cite{Piilo2008}. However, the extension of the state space of an open system may retrieve this key form~\cite{Breuer2004,Hush2015}. The physical relevance of our results needs to be clarified precisely. Finally, the role of many-body interactions of the quantum system in stochastic thermodynamics is almost unexplored issue; QJT would be a basic and useful notion to investigate this issue~\cite{Daley2014}.
\\
\\
{\noindent \it Acknowledgment.} We would like to thank Editage (http://www.editage.com) for English language editing.
The work was supported by the National Science Foundation of China under Grant Nos. 11174025 and 11575016.

\section*{Appendix A: Multitime correlation function of operators}
There is a very general definition for the multitime correlation function~\cite{Breuer2002,Gardiner2004}. Here, we restrict it to the following form that is relevant to our applications:
\begin{eqnarray}
\label{Multitimecorrelationfunction}
&&{\rm Tr}\left[ O_N(t_N)G(t_N,t_{N-1})( \cdots O_1(t_1) G(t_1,t_0)[\rho(t_0)]B_1(t_1)  \cdots )B_N(t_N)\right]\nonumber\\
&=&\left\langle O_N(t_N)\cdots O_1(t_1)B_1(t_1)\cdots B_N(t_N)\right \rangle,
\end{eqnarray}
where $O_i$ and $B_i$, $i=1,\cdots,N$, are arbitrary operators, and $t_i$ are ordered in time such that $t_N> \cdots >t_1> t_0$.

\section*{Appendix B: CFs using multitime correlation function of operators}
The structure of Eq.~(\ref{workmasterequationforward}) is the same as that of Eq.~(\ref{peturbedmastereqheat}). Hence, we can express the CF of the work in Eq.~(\ref{CFworkalternativeform}) by using the multitime correlation function of operators. To this end, we rewrite the time-evolution equation of the operator $K(t,t_0;\eta)$,
\begin{eqnarray}
\label{workmasterequationforward1}
\partial_t K&=&{\cal L}(t)K+\partial_t e^{i\eta H'(t)}e^{-i\eta H'(t)}K-\frac{i}{\hbar}[e^{i\eta H'(t)},H(t)]e^{-i\eta H'(t)}K,\nonumber\\
&=&{\cal L}(t)K+{\cal W}_\eta(t)K,
\end{eqnarray}
and its initial condition is $\rho'(t_0)$, where the action of ${\cal W}_\eta$ on an operator is a simple multiplication from the right-hand of the operator. Applying the Dyson's series again, we obtain the following alternative formal solution:
\begin{eqnarray}
K(t,t_0;\eta)= \int_C {\cal D}(t)\hspace{0.1cm} G(t,t_N){\cal W}_{\eta}(t_N)G(t_N,t_{N-1})\cdots {\cal W}_{\eta}(t_1)G(t_1,t_0)[\rho'(t_0)].
\end{eqnarray}
Using the definition~(\ref{Multitimecorrelationfunction}), we immediately find a concise expression of the CF of the work:
\begin{eqnarray}
\label{workequalityFKform}
\Psi(\eta)&=&\left\langle {\cal T}_{\leftarrow}e^{\int_{t_0}^{t} d\tau {\cal W}_\eta(\tau)}\right \rangle'.
\end{eqnarray}
Although this new form does not bring any advantages in computing over the original ones, Eqs.~(\ref{CFworkfinal}) and~(\ref{CFworkalternativeform}), it indeed corresponds to the celebrated Feynman-Kac formula in classical stochastic processes~\cite{Kac1949,Chetrite2012,Liu2012,Liu2014,Liu2014a,Liu2014b}. Analogously, the concise expressions of the CFs of the heat and entropy production are given by
\begin{eqnarray}
\Phi(\xi)&=&\left\langle e^{-i\xi H'(t_f)}{\cal T}_{\leftarrow}e^{\int_{t_0}^{t} d\tau  {\cal W}_\xi(\tau)}e^{i\xi H'(t_0)}\right\rangle,\\
\Omega(\zeta)&=&\left\langle \rho(t_f)^{-i\zeta k_B}e^{-i\zeta H'(t_f)/T}{\cal T}_{\leftarrow}e^{\int_{t_0}^{t} d\tau  {\cal W}_{\zeta/T}(\tau)}e^{i\zeta H'(t_0)/T}\rho(t_0)^{i\zeta k_B}\right\rangle,
\end{eqnarray}
respectively. The last equation with specific $\zeta=i/k_B$ has been found earlier by one of the authors~\cite{Liu2012a}. However, its physical explanation is clarified only here.

\section*{Appendix C: Backward-time-evolution equations}

In order to derive the backward time counterpart of Eq.~(\ref{workmasterequationforward}), it is convenient to apply the following property:
\begin{eqnarray}
{\rm Tr}[O_1\hat G(t,t';\eta)(O_2)]&=&{\rm Tr}[{\hat G}^\star (t',t;\eta)(O_1) O_2],
\end{eqnarray}
where the two times $t\ge t'$; $O_i$, $i=1,2$, are two arbitrary operators; and the superpropgator $\hat{G}^\star(t',t;\eta)$ is equal to ${\cal T}_\rightarrow \exp[{\int_{t'}^t d\tau\hat{\cal{L}}^\star(\tau;\eta)}]$  with
\begin{eqnarray}
\hat{\cal {L}}^\star(t;\eta)(O_1)&=&\frac{i}{\hbar}[H(t),O_1]+\sum_{\omega_t}\gamma(\omega_t)\left[A^\dag (\omega_t,t)O_1 A(\omega_t,t)-\frac{1}{2}\left\{ A^\dag(\omega_t,t)A(\omega_t,t),O_1\right \}\right]\nonumber\\
&+&O_1\partial_t e^{i\eta H'(t)}e^{-i\eta H'(t)}+O_1\frac{i}{\hbar}[H(t),e^{i\eta H'(t)}]e^{-i\eta H'(t)},
\end{eqnarray}
where ${\cal T}_\rightarrow $ denotes the antichronological time-ordering operator.
The proof is direct, and we will not show it here. On the basis of this property, we can immediately rewrite Eq.~(\ref{CFworkalternativeform}) as
\begin{eqnarray}
\Psi(\eta)={\rm Tr}[\hat G^\star(t_0,t;\eta)(I)\rho'(t_0)]={\rm Tr}[K^\star(t_0,t;\eta)\rho'(t_0)],
\end{eqnarray}
where the time-evolution equation of the new operator $K^\star(t',t;\eta)$ with respect to the backward time $t'$ is
\begin{eqnarray}
\label{workmasterequationbackward}
\partial_{t'} K^\star=-\hat{{\cal L}}^\star(t';\eta)K^\star,
\end{eqnarray}
and its terminal rather than initial condition is $K(t,t;\eta)=I$. Two specific cases of Eq.~(\ref{workmasterequationbackward}) have been given in Refs.~\cite{Liu2014} and~\cite{Liu2014a}.

\section*{Appendix D: Distributions of heat of a Floquet QME}
\subsection{CF method}
Because the Floquet basis, Eq.~(\ref{Floquetbasis}), is complete and orthogonal, it is convenient to expand Eq.~(\ref{heatmasterequation}) in this basis. First, the Lindblad operators~(\ref{Lindbladoperators}) are as follows:
\begin{eqnarray}
A(\omega_L,t)&\doteq&\frac{\Omega}{2\Omega'}\sigma_z(t),\nonumber\\
A(\omega_L-\Omega',t)&\doteq&\left(\frac{\delta-\Omega'}{2\Omega'}\right)\sigma_+(t),\nonumber\\
A(\omega_L+\Omega',t)&\doteq&\left(\frac{\delta+\Omega'}{2\Omega'}\right)\sigma_-(t) ,
\end{eqnarray}
and the other three Lindblad operators $A(\omega,t)$ with $\omega=-\omega_L$, $-(\omega_L-\Omega')$, and $-(\omega_L+\Omega')$ are their adjoint operators. Note that, in order to indicate that these Pauli matrixes are not the conventional ones, we add time parameters after these symbols. Using these matrixes, we expand the operator $\hat{{\rho}}(t,t_0;\xi)$ as follows:
\begin{eqnarray}
\hat{\rho}=\frac{p_+(t)+p_-(t)}{2}I+\frac{p_+(t)-p_-(t)}{2}\sigma_z(t)+p_1(t)\sigma_+(t)+p_2(t)\sigma_-(t),
\end{eqnarray}
where $p_{\pm}$ and $p_{i}$, $i=1,2$, are the diagonal and non-diagonal elements of this density matrix. Substituting them into Eq.~(\ref{heatmasterequation}) and doing a simple algebra, we get the time-evolution equations for $\hat{p}_\pm(t)$:
\begin{eqnarray}
\frac{dp_\pm}{dt}&=&\left[(e^{i\xi\omega_L}-1)\gamma(\omega_L)\left(\frac{\Omega}{2\Omega'}\right)^2 +    (e^{-i\xi\omega_L}-1)\gamma(-\omega_L)\left(\frac{\Omega}{2\Omega'}\right)^2\right. \nonumber\\
&&-\left.\gamma(\mp(\omega_L-\Omega'))\left(\frac{\delta-\Omega'}{2\Omega'}\right)^2-
\gamma(\pm(\omega_L+\Omega'))\left(\frac{\delta+\Omega'}{2\Omega'}\right)^2\right]p_\pm\nonumber\\
&&+\left[e^{\pm i\xi(\omega_L-\Omega')}\gamma(\pm(\omega_L-\Omega'))\left(\frac{\delta-\Omega'}{2\Omega'}\right)^2
    +e^{\mp i\xi(\omega_L+\Omega')}\gamma(\mp(\omega_L+\Omega'))\left(\frac{\delta+\Omega'}{2\Omega'}\right)^2
\right]p_\mp.
\end{eqnarray}
Because the initial density matrix is assumed to be the identity operator, the initial conditions of $p_{\pm}$ are simply $1/2$. Although these two equations are a bit long, they are the first-order ordinary differential equations with constant coefficients. Their solutions are simple. Obviously, the CF of the heat is $\Phi(\xi)=p_+(t)+p_-(t)$. By substituting the concrete rates and performing an inverse Fourier transform, we can obtain the distribution of the heat. What we did here is completely parallel with those done in Ref.~\cite{Gasparinetti2014}.

\subsection{Simulation of QJTs}
According to the notion of QJT~\cite{Plenio1998,Breuer2002,Carmichael1993,Wiseman2010}, the reduced density matrix $\rho(t)$ of the QME~(\ref{masterequation}) can be interpreted as a statistical average of wave vectors. The wave vector $\Psi(t)$ varies in the Hilbert space of the TLS with alternatively deterministic continuous evolution and stochastic jumps. Assuming that the continuous evolution starts from time $t$ and ends at time $t+\tau$, during this process its deterministic equation is
\begin{eqnarray}
\label{wavevectorevolvingeq}
\frac{d}{ds}\Psi(t+s)=-\frac{i}{\hbar} \left[H(t+s)-\frac{i\hbar}{2}\sum_{\omega_{t+s}}\gamma(\omega_{t+s}),  A^\dag(\omega_{t+s},t+s)A(\omega_{t+s},t+s)\right]\Psi(t+s),
\end{eqnarray}
$0\le s\le \tau$. Because the Floquet basis is complete and orthogonal, it is convenient to expand $\Psi(t+s)$ in this basis, that is,
\begin{eqnarray}
\label{wavevectorunnormalized}
\Psi(t+s)=\mu_+(s)|u_+(t+s)\rangle +\mu_-(s)|u_-(t+s)\rangle.
\end{eqnarray}
Substituting it into Eq.~(\ref{wavevectorevolvingeq}), we get
\begin{eqnarray}
\label{componentsequation1}
  \frac{d\mu_\pm}{ds}&=&-i\frac{\varepsilon_\pm}{\hbar}\mu_\pm-\frac{1}{\tau_\pm}\mu_\pm,
\end{eqnarray}
where the coefficients are
\begin{eqnarray}
\frac{1}{\tau_\pm}=\frac{1}{2}\left[\left(\frac{\Omega}{2\Omega'}\right)^2\left(
    \gamma(\omega_L) +
    \gamma(-\omega_L) \right)+\gamma(\mp(\omega_L-\Omega'))\left(\frac{\delta-\Omega'}{2\Omega'}\right)^2
    +\gamma(\pm(\omega_L+\Omega'))\left(\frac{\delta+\Omega'}{2\Omega'}\right)^2\right],
\end{eqnarray}
respectively. Eqs.~(\ref{componentsequation1}) have simple solutions,
\begin{eqnarray}
\mu_\pm(s)&=&\mu_\pm(0)\exp\left[-\left(i\frac{\varepsilon_\pm}{\hbar}+\frac{1}{\tau_\pm}\right)s\right].
\end{eqnarray}
Obviously, the wave vector, $\Psi(t+s)$, is not normalized. The normalized one is $\overline{\Psi}(t+s)$, which is the same as Eq.~(\ref{wavevectorunnormalized}) except that $\mu_\pm(s)$ therein are replaced by $\overline{\mu}_\pm(s)=\mu_\pm(s)/\sqrt{\|\mu_+(s)\|^2+\|\mu_-(s)\|^2} $. We can determine the time duration $\tau$ by solving equation
\begin{eqnarray}
\eta =\|\Psi(t+\tau)\|^2= \|\mu_+(0)\|^2\exp\left(-\frac{2\tau}{\tau_+}\right)+\|\mu_-(0)\|^2\exp\left(-\frac{2\tau}{\tau_-}\right),
\end{eqnarray}
where $\eta\in(0,1)$ is an uniform random number.

This smooth evolution is interrupted by a jump at time $t+\tau$. The state after the jump is
\begin{eqnarray}
A(\omega_{t+\tau},t+\tau)\Psi(t+\tau)/\| A(\omega_{t+\tau},t+\tau)\Psi(t+\tau)\|.
\end{eqnarray}
The probabilities of these jumps are proportional to
\begin{eqnarray}
\gamma(\omega_{t+\tau})\|A(\omega_{t+\tau},t+\tau)\Psi(t+\tau)\|^2.
\end{eqnarray}
We list these six states in the following table:
\begin{eqnarray}
\begin{array}{ccccccccccccccccc}
  \hbox{State after a jump} && \hbox{Probabilities }\propto && \hbox{Heat produced }\\
  \bar{\mu}_+(\tau)|u_+(t+\tau)\rangle-\bar{\mu}_-(\tau)|u_-(t+\tau)\rangle && \gamma(\omega_L)({\Omega}/{2\Omega'})^2 && \hbar\omega_L\\
  |u_+(t+\tau)\rangle \hspace{0.5cm}({\rm if}\hspace{0.2cm} \mu_-\neq0)&& \gamma(\omega_L-\Omega')(\|\mu_-(\tau)\|(\delta-\Omega')/{2\Omega'})^2&& \hbar(\omega_L-\Omega')\\
  |u_-(t+\tau)\rangle \hspace{0.5cm}({\rm if}\hspace{0.2cm}  \mu_+\neq0)&& \gamma(\omega_L+\Omega')((\|\mu_+(\tau)\|\delta+\Omega')/{2\Omega'})^2 &&\hbar(\omega_L+\Omega')\\
    \bar{\mu}_+(\tau)|u_+(t+\tau)\rangle-\bar{\mu}_-(\tau)|u_-(t+\tau)\rangle && \gamma(\omega_L)({\Omega}/{2\Omega'})^2 &&-\hbar\omega_L\\
  |u_-(t+\tau)\rangle \hspace{0.5cm}({\rm if}\hspace{0.2cm}  \mu_+\neq0)&& \gamma(-(\omega_L-\Omega'))(\|\mu_+(\tau)\|(\delta-\Omega')/{2\Omega'})^2 && -\hbar(\omega_L-\Omega')\\
  |u_+(t+\tau)\rangle \hspace{0.5cm}({\rm if}\hspace{0.2cm}  \mu_-\neq0)&&  \gamma(-(\omega_L+\Omega'))(\|\mu_-(\tau)\|(\delta+\Omega')/{2\Omega'})^2 &&-\hbar(\omega_L+\Omega')
  \end{array}
 \end{eqnarray}
After a state is randomly chosen from them, new rounds with continuous evolution and stochastic jump start until the end time is arrived.


\begin{thebibliography}{111}%
\makeatletter
\providecommand \@ifxundefined [1]{%
 \@ifx{#1\undefined}
}%
\providecommand \@ifnum [1]{%
 \ifnum #1\expandafter \@firstoftwo
 \else \expandafter \@secondoftwo
 \fi
}%
\providecommand \@ifx [1]{%
 \ifx #1\expandafter \@firstoftwo
 \else \expandafter \@secondoftwo
 \fi
}%
\providecommand \natexlab [1]{#1}%
\providecommand \enquote  [1]{``#1''}%
\providecommand \bibnamefont  [1]{#1}%
\providecommand \bibfnamefont [1]{#1}%
\providecommand \citenamefont [1]{#1}%
\providecommand \href@noop [0]{\@secondoftwo}%
\providecommand \href [0]{\begingroup \@sanitize@url \@href}%
\providecommand \@href[1]{\@@startlink{#1}\@@href}%
\providecommand \@@href[1]{\endgroup#1\@@endlink}%
\providecommand \@sanitize@url [0]{\catcode `\\12\catcode `\$12\catcode
  `\&12\catcode `\#12\catcode `\^12\catcode `\_12\catcode `\%12\relax}%
\providecommand \@@startlink[1]{}%
\providecommand \@@endlink[0]{}%
\providecommand \url  [0]{\begingroup\@sanitize@url \@url }%
\providecommand \@url [1]{\endgroup\@href {#1}{\urlprefix }}%
\providecommand \urlprefix  [0]{URL }%
\providecommand \Eprint [0]{\href }%
\providecommand \doibase [0]{http://dx.doi.org/}%
\providecommand \selectlanguage [0]{\@gobble}%
\providecommand \bibinfo  [0]{\@secondoftwo}%
\providecommand \bibfield  [0]{\@secondoftwo}%
\providecommand \translation [1]{[#1]}%
\providecommand \BibitemOpen [0]{}%
\providecommand \bibitemStop [0]{}%
\providecommand \bibitemNoStop [0]{.\EOS\space}%
\providecommand \EOS [0]{\spacefactor3000\relax}%
\providecommand \BibitemShut  [1]{\csname bibitem#1\endcsname}%
\let\auto@bib@innerbib\@empty
\bibitem [{\citenamefont {Mukamel}(2003)}]{Mukamel2003}%
  \BibitemOpen
  \bibfield  {author} {\bibinfo {author} {\bibfnamefont {S.}~\bibnamefont
  {Mukamel}},\ }\href {\doibase 10.1103/PhysRevLett.90.170604} {\bibfield
  {journal} {\bibinfo  {journal} {Phys. Rev. Lett.}\ }\textbf {\bibinfo
  {volume} {90}},\ \bibinfo {pages} {170604} (\bibinfo {year}
  {2003})}\BibitemShut {NoStop}%
\bibitem [{\citenamefont {De~Roeck}\ and\ \citenamefont
  {Maes}(2004)}]{DeRoeck2004}%
  \BibitemOpen
  \bibfield  {author} {\bibinfo {author} {\bibfnamefont {W.}~\bibnamefont
  {De~Roeck}}\ and\ \bibinfo {author} {\bibfnamefont {C.}~\bibnamefont
  {Maes}},\ }\href {http://pre.aps.org/abstract/PRE/v69/i2/e026115} {\bibfield
  {journal} {\bibinfo  {journal} {Phys. Rev. E}\ }\textbf {\bibinfo {volume}
  {69}},\ \bibinfo {pages} {026115} (\bibinfo {year} {2004})}\BibitemShut
  {NoStop}%
\bibitem [{\citenamefont {De~Roeck}(2007)}]{DeRoeck2007}%
  \BibitemOpen
  \bibfield  {author} {\bibinfo {author} {\bibfnamefont {W.}~\bibnamefont
  {De~Roeck}},\ }\href
  {http://www.sciencedirect.com/science/article/pii/S1631070507000953}
  {\bibfield  {journal} {\bibinfo  {journal} {C. R. Phys.}\ }\textbf {\bibinfo
  {volume} {8}},\ \bibinfo {pages} {674} (\bibinfo {year} {2007})}\BibitemShut
  {NoStop}%
\bibitem [{\citenamefont {Esposito}\ \emph {et~al.}(2009)\citenamefont
  {Esposito}, \citenamefont {Harbola},\ and\ \citenamefont
  {Mukamel}}]{Esposito2009}%
  \BibitemOpen
  \bibfield  {author} {\bibinfo {author} {\bibfnamefont {M.}~\bibnamefont
  {Esposito}}, \bibinfo {author} {\bibfnamefont {U.}~\bibnamefont {Harbola}}, \
  and\ \bibinfo {author} {\bibfnamefont {S.}~\bibnamefont {Mukamel}},\ }\href
  {http://rmp.aps.org/abstract/RMP/v81/i4/p1665_1} {\bibfield  {journal}
  {\bibinfo  {journal} {Rev. Mod. Phys.}\ }\textbf {\bibinfo {volume} {81}},\
  \bibinfo {pages} {1665} (\bibinfo {year} {2009})}\BibitemShut {NoStop}%
\bibitem [{\citenamefont {Horowitz}\ and\ \citenamefont
  {Parrondo}(2013)}]{Horowitz2013}%
  \BibitemOpen
  \bibfield  {author} {\bibinfo {author} {\bibfnamefont {J.~M.}\ \bibnamefont
  {Horowitz}}\ and\ \bibinfo {author} {\bibfnamefont {J.~M.}\ \bibnamefont
  {Parrondo}},\ }\href {http://iopscience.iop.org/1367-2630/15/8/085028}
  {\bibfield  {journal} {\bibinfo  {journal} {New J. Phys.}\ }\textbf {\bibinfo
  {volume} {15}},\ \bibinfo {pages} {085028} (\bibinfo {year}
  {2013})}\BibitemShut {NoStop}%
\bibitem [{\citenamefont {Esposito}\ and\ \citenamefont
  {Mukamel}(2006)}]{Esposito2006}%
  \BibitemOpen
  \bibfield  {author} {\bibinfo {author} {\bibfnamefont {M.}~\bibnamefont
  {Esposito}}\ and\ \bibinfo {author} {\bibfnamefont {S.}~\bibnamefont
  {Mukamel}},\ }\href {http://pre.aps.org/abstract/PRE/v73/i4/e046129}
  {\bibfield  {journal} {\bibinfo  {journal} {Phys. Rev. E}\ }\textbf {\bibinfo
  {volume} {73}},\ \bibinfo {pages} {046129} (\bibinfo {year}
  {2006})}\BibitemShut {NoStop}%
\bibitem [{\citenamefont {Crooks}(2008)}]{Crooks2008}%
  \BibitemOpen
  \bibfield  {author} {\bibinfo {author} {\bibfnamefont {G.~E.}\ \bibnamefont
  {Crooks}},\ }\href {\doibase 10.1103/PhysRevA.77.034101} {\bibfield
  {journal} {\bibinfo  {journal} {Phys. Rev. A}\ }\textbf {\bibinfo {volume}
  {77}},\ \bibinfo {pages} {034101} (\bibinfo {year} {2008})}\BibitemShut
  {NoStop}%
\bibitem [{\citenamefont {Talkner}\ \emph {et~al.}(2009)\citenamefont
  {Talkner}, \citenamefont {Campisi},\ and\ \citenamefont
  {H{\"a}nggi}}]{Talkner2009}%
  \BibitemOpen
  \bibfield  {author} {\bibinfo {author} {\bibfnamefont {P.}~\bibnamefont
  {Talkner}}, \bibinfo {author} {\bibfnamefont {M.}~\bibnamefont {Campisi}}, \
  and\ \bibinfo {author} {\bibfnamefont {P.}~\bibnamefont {H{\"a}nggi}},\
  }\href {http://iopscience.iop.org/1742-5468/2009/02/P02025} {\bibfield
  {journal} {\bibinfo  {journal} {J. Stat. Mech.: Theor. Exp.}\ }\textbf
  {\bibinfo {volume} {2009}},\ \bibinfo {pages} {P02025} (\bibinfo {year}
  {2009})}\BibitemShut {NoStop}%
\bibitem [{\citenamefont {Horowitz}(2012)}]{Horowitz2012}%
  \BibitemOpen
  \bibfield  {author} {\bibinfo {author} {\bibfnamefont {J.~M.}\ \bibnamefont
  {Horowitz}},\ }\href {http://pre.aps.org/abstract/PRE/v85/i3/e031110}
  {\bibfield  {journal} {\bibinfo  {journal} {Phys. Rev. E}\ }\textbf {\bibinfo
  {volume} {85}},\ \bibinfo {pages} {031110} (\bibinfo {year}
  {2012})}\BibitemShut {NoStop}%
\bibitem [{\citenamefont {Liu}(2012{\natexlab{a}})}]{Liu2012}%
  \BibitemOpen
  \bibfield  {author} {\bibinfo {author} {\bibfnamefont {F.}~\bibnamefont
  {Liu}},\ }\href {\doibase 10.1103/PhysRevE.86.010103} {\bibfield  {journal}
  {\bibinfo  {journal} {Phys. Rev. E}\ }\textbf {\bibinfo {volume} {86}},\
  \bibinfo {pages} {010103} (\bibinfo {year} {2012}{\natexlab{a}})}\BibitemShut
  {NoStop}%
\bibitem [{\citenamefont {Chetrite}\ and\ \citenamefont
  {Mallick}(2012)}]{Chetrite2012}%
  \BibitemOpen
  \bibfield  {author} {\bibinfo {author} {\bibfnamefont {R.}~\bibnamefont
  {Chetrite}}\ and\ \bibinfo {author} {\bibfnamefont {K.}~\bibnamefont
  {Mallick}},\ }\href
  {http://link.springer.com/article/10.1007/s10955-012-0557-z} {\bibfield
  {journal} {\bibinfo  {journal} {J. Stat. Phys.}\ }\textbf {\bibinfo {volume}
  {148}},\ \bibinfo {pages} {480} (\bibinfo {year} {2012})}\BibitemShut
  {NoStop}%
\bibitem [{\citenamefont {Jaksic}\ \emph {et~al.}(2013)\citenamefont {Jaksic},
  \citenamefont {Pillet},\ and\ \citenamefont {Westrich}}]{Jaksic2013}%
  \BibitemOpen
  \bibfield  {author} {\bibinfo {author} {\bibfnamefont {V.}~\bibnamefont
  {Jaksic}}, \bibinfo {author} {\bibfnamefont {C.~A.}\ \bibnamefont {Pillet}},
  \ and\ \bibinfo {author} {\bibfnamefont {M.}~\bibnamefont {Westrich}},\
  }\href@noop {} {\bibfield  {journal} {\bibinfo  {journal} {J. Stat. Phys.}\
  }\textbf {\bibinfo {volume} {154}},\ \bibinfo {pages} {153} (\bibinfo {year}
  {2013})}\BibitemShut {NoStop}%
\bibitem [{\citenamefont {Hekking}\ and\ \citenamefont
  {Pekola}(2013)}]{Hekking2013}%
  \BibitemOpen
  \bibfield  {author} {\bibinfo {author} {\bibfnamefont {F.~W.~J.}\
  \bibnamefont {Hekking}}\ and\ \bibinfo {author} {\bibfnamefont {J.~P.}\
  \bibnamefont {Pekola}},\ }\href
  {http://prl.aps.org/abstract/PRL/v111/i9/e093602} {\bibfield  {journal}
  {\bibinfo  {journal} {Phys. Rev. Lett.}\ }\textbf {\bibinfo {volume} {111}},\
  \bibinfo {pages} {093602} (\bibinfo {year} {2013})}\BibitemShut {NoStop}%
\bibitem [{\citenamefont {Leggio}\ \emph {et~al.}(2013)\citenamefont {Leggio},
  \citenamefont {Napoli}, \citenamefont {Messina},\ and\ \citenamefont
  {Breuer}}]{Leggio2013}%
  \BibitemOpen
  \bibfield  {author} {\bibinfo {author} {\bibfnamefont {B.}~\bibnamefont
  {Leggio}}, \bibinfo {author} {\bibfnamefont {A.}~\bibnamefont {Napoli}},
  \bibinfo {author} {\bibfnamefont {A.}~\bibnamefont {Messina}}, \ and\
  \bibinfo {author} {\bibfnamefont {H.-P.}\ \bibnamefont {Breuer}},\ }\href
  {\doibase 10.1103/PhysRevA.88.042111} {\bibfield  {journal} {\bibinfo
  {journal} {Phys. Rev. A}\ }\textbf {\bibinfo {volume} {88}},\ \bibinfo
  {pages} {042111} (\bibinfo {year} {2013})}\BibitemShut {NoStop}%
\bibitem [{\citenamefont {Liu}(2014{\natexlab{a}})}]{Liu2014}%
  \BibitemOpen
  \bibfield  {author} {\bibinfo {author} {\bibfnamefont {F.}~\bibnamefont
  {Liu}},\ }\href {\doibase 10.1103/PhysRevE.89.042122} {\bibfield  {journal}
  {\bibinfo  {journal} {Phys. Rev. E}\ }\textbf {\bibinfo {volume} {89}},\
  \bibinfo {pages} {042122} (\bibinfo {year} {2014}{\natexlab{a}})}\BibitemShut
  {NoStop}%
\bibitem [{\citenamefont {Liu}(2014{\natexlab{b}})}]{Liu2014a}%
  \BibitemOpen
  \bibfield  {author} {\bibinfo {author} {\bibfnamefont {F.}~\bibnamefont
  {Liu}},\ }\href@noop {} {\bibfield  {journal} {\bibinfo  {journal} {Phys.
  Rev. E}\ }\textbf {\bibinfo {volume} {90}},\ \bibinfo {pages} {032121}
  (\bibinfo {year} {2014}{\natexlab{b}})}\BibitemShut {NoStop}%
\bibitem [{\citenamefont {Silaev}\ \emph {et~al.}(2014)\citenamefont {Silaev},
  \citenamefont {Heikkil\"a},\ and\ \citenamefont {Virtanen}}]{Silaev2014}%
  \BibitemOpen
  \bibfield  {author} {\bibinfo {author} {\bibfnamefont {M.}~\bibnamefont
  {Silaev}}, \bibinfo {author} {\bibfnamefont {T.~T.}\ \bibnamefont
  {Heikkil\"a}}, \ and\ \bibinfo {author} {\bibfnamefont {P.}~\bibnamefont
  {Virtanen}},\ }\href {\doibase 10.1103/PhysRevE.90.022103} {\bibfield
  {journal} {\bibinfo  {journal} {Phys. Rev. E}\ }\textbf {\bibinfo {volume}
  {90}},\ \bibinfo {pages} {022103} (\bibinfo {year} {2014})}\BibitemShut
  {NoStop}%
\bibitem [{\citenamefont {Suomela}\ \emph {et~al.}(2014)\citenamefont
  {Suomela}, \citenamefont {Solinas}, \citenamefont {Pekola}, \citenamefont
  {Ankerhold},\ and\ \citenamefont {Ala-Nissila}}]{Suomela2014}%
  \BibitemOpen
  \bibfield  {author} {\bibinfo {author} {\bibfnamefont {S.}~\bibnamefont
  {Suomela}}, \bibinfo {author} {\bibfnamefont {P.}~\bibnamefont {Solinas}},
  \bibinfo {author} {\bibfnamefont {J.~P.}\ \bibnamefont {Pekola}}, \bibinfo
  {author} {\bibfnamefont {J.}~\bibnamefont {Ankerhold}}, \ and\ \bibinfo
  {author} {\bibfnamefont {T.}~\bibnamefont {Ala-Nissila}},\ }\href {\doibase
  10.1103/PhysRevB.90.094304} {\bibfield  {journal} {\bibinfo  {journal} {Phys.
  Rev. B}\ }\textbf {\bibinfo {volume} {90}},\ \bibinfo {pages} {094304}
  (\bibinfo {year} {2014})}\BibitemShut {NoStop}%
\bibitem [{\citenamefont {Suomela}\ \emph {et~al.}(2015)\citenamefont
  {Suomela}, \citenamefont {Salmilehto}, \citenamefont {Savenko}, \citenamefont
  {Ala-Nissila},\ and\ \citenamefont {M{\"{o}}tt{\"{o}}nen}}]{Suomela2015}%
  \BibitemOpen
  \bibfield  {author} {\bibinfo {author} {\bibfnamefont {S.}~\bibnamefont
  {Suomela}}, \bibinfo {author} {\bibfnamefont {J.}~\bibnamefont {Salmilehto}},
  \bibinfo {author} {\bibfnamefont {I.~G.}\ \bibnamefont {Savenko}}, \bibinfo
  {author} {\bibfnamefont {T.}~\bibnamefont {Ala-Nissila}}, \ and\ \bibinfo
  {author} {\bibfnamefont {M.}~\bibnamefont {M{\"{o}}tt{\"{o}}nen}},\
  }\href@noop {} {\bibfield  {journal} {\bibinfo  {journal} {Phys. Rev. E}\
  }\textbf {\bibinfo {volume} {91}},\ \bibinfo {pages} {022126} (\bibinfo
  {year} {2015})}\BibitemShut {NoStop}%
\bibitem [{\citenamefont {Manzano}\ \emph {et~al.}(2015)\citenamefont
  {Manzano}, \citenamefont {Horowitz},\ and\ \citenamefont
  {Parrondo}}]{Manzano2015}%
  \BibitemOpen
  \bibfield  {author} {\bibinfo {author} {\bibfnamefont {G.}~\bibnamefont
  {Manzano}}, \bibinfo {author} {\bibfnamefont {J.~M.}\ \bibnamefont
  {Horowitz}}, \ and\ \bibinfo {author} {\bibfnamefont {J.~M.~R.}\ \bibnamefont
  {Parrondo}},\ }\href {\doibase 10.1103/PhysRevE.92.032129} {\bibfield
  {journal} {\bibinfo  {journal} {Phys. Rev. E}\ }\textbf {\bibinfo {volume}
  {92}},\ \bibinfo {pages} {032129} (\bibinfo {year} {2015})}\BibitemShut
  {NoStop}%
\bibitem [{\citenamefont {Gasparinetti}\ \emph {et~al.}(2014)\citenamefont
  {Gasparinetti}, \citenamefont {Solinas}, \citenamefont {Braggio},\ and\
  \citenamefont {Sassetti}}]{Gasparinetti2014}%
  \BibitemOpen
  \bibfield  {author} {\bibinfo {author} {\bibfnamefont {S.}~\bibnamefont
  {Gasparinetti}}, \bibinfo {author} {\bibfnamefont {P.}~\bibnamefont
  {Solinas}}, \bibinfo {author} {\bibfnamefont {A.}~\bibnamefont {Braggio}}, \
  and\ \bibinfo {author} {\bibfnamefont {M.}~\bibnamefont {Sassetti}},\ }\href
  {http://stacks.iop.org/1367-2630/16/i=11/a=115001} {\bibfield  {journal}
  {\bibinfo  {journal} {New J. Phys.}\ }\textbf {\bibinfo {volume} {16}},\
  \bibinfo {pages} {115001} (\bibinfo {year} {2014})}\BibitemShut {NoStop}%
\bibitem [{\citenamefont {Cuetara}\ \emph {et~al.}(2015)\citenamefont
  {Cuetara}, \citenamefont {Engel},\ and\ \citenamefont
  {Esposito}}]{Cuetara2015}%
  \BibitemOpen
  \bibfield  {author} {\bibinfo {author} {\bibfnamefont {G.~B.}\ \bibnamefont
  {Cuetara}}, \bibinfo {author} {\bibfnamefont {A.}~\bibnamefont {Engel}}, \
  and\ \bibinfo {author} {\bibfnamefont {M.}~\bibnamefont {Esposito}},\
  }\href@noop {} {\bibfield  {journal} {\bibinfo  {journal} {New J. Phys.}\
  }\textbf {\bibinfo {volume} {17}},\ \bibinfo {pages} {055002} (\bibinfo
  {year} {2015})}\BibitemShut {NoStop}%
\bibitem [{\citenamefont {Alonso}\ \emph {et~al.}(2016)\citenamefont {Alonso},
  \citenamefont {Lutz},\ and\ \citenamefont {Romito}}]{Alonso2016}%
  \BibitemOpen
  \bibfield  {author} {\bibinfo {author} {\bibfnamefont {J.~J.}\ \bibnamefont
  {Alonso}}, \bibinfo {author} {\bibfnamefont {E.}~\bibnamefont {Lutz}}, \ and\
  \bibinfo {author} {\bibfnamefont {A.}~\bibnamefont {Romito}},\ }\href
  {\doibase 10.1103/PhysRevLett.116.080403} {\bibfield  {journal} {\bibinfo
  {journal} {Phys. Rev. Lett.}\ }\textbf {\bibinfo {volume} {116}},\ \bibinfo
  {pages} {080403} (\bibinfo {year} {2016})}\BibitemShut {NoStop}%
\bibitem [{\citenamefont {Gong}\ \emph {et~al.}(2016)\citenamefont {Gong},
  \citenamefont {Ashida},\ and\ \citenamefont {Ueda}}]{Gong2016}%
  \BibitemOpen
  \bibfield  {author} {\bibinfo {author} {\bibfnamefont {Z.}~\bibnamefont
  {Gong}}, \bibinfo {author} {\bibfnamefont {Y.}~\bibnamefont {Ashida}}, \ and\
  \bibinfo {author} {\bibfnamefont {M.}~\bibnamefont {Ueda}},\ }\href {\doibase
  10.1103/PhysRevA.94.012107} {\bibfield  {journal} {\bibinfo  {journal} {Phys.
  Rev. A}\ }\textbf {\bibinfo {volume} {94}},\ \bibinfo {pages} {012107}
  (\bibinfo {year} {2016})}\BibitemShut {NoStop}%
\bibitem [{\citenamefont {Davies}(1974)}]{Davies1974}%
  \BibitemOpen
  \bibfield  {author} {\bibinfo {author} {\bibfnamefont {E.~B.}\ \bibnamefont
  {Davies}},\ }\href {http://link.springer.com/article/10.1007/BF01608389}
  {\bibfield  {journal} {\bibinfo  {journal} {Comm. Math. Phys.}\ }\textbf
  {\bibinfo {volume} {39}},\ \bibinfo {pages} {91} (\bibinfo {year}
  {1974})}\BibitemShut {NoStop}%
\bibitem [{\citenamefont {Lindblad}(1976)}]{Lindblad1976}%
  \BibitemOpen
  \bibfield  {author} {\bibinfo {author} {\bibfnamefont {G.}~\bibnamefont
  {Lindblad}},\ }\href {http://link.springer.com/article/10.1007/BF01608499}
  {\bibfield  {journal} {\bibinfo  {journal} {Comm. Math. Phys.}\ }\textbf
  {\bibinfo {volume} {48}},\ \bibinfo {pages} {119} (\bibinfo {year}
  {1976})}\BibitemShut {NoStop}%
\bibitem [{\citenamefont {Gorini}\ \emph {et~al.}(1976)\citenamefont {Gorini},
  \citenamefont {Kossakowski},\ and\ \citenamefont {Sudarshan}}]{Gorini1976}%
  \BibitemOpen
  \bibfield  {author} {\bibinfo {author} {\bibfnamefont {V.}~\bibnamefont
  {Gorini}}, \bibinfo {author} {\bibfnamefont {A.}~\bibnamefont {Kossakowski}},
  \ and\ \bibinfo {author} {\bibfnamefont {E.~C.~G.}\ \bibnamefont
  {Sudarshan}},\ }\href@noop {} {\bibfield  {journal} {\bibinfo  {journal} {J.
  Math. Phys.}\ }\textbf {\bibinfo {volume} {17}},\ \bibinfo {pages} {821}
  (\bibinfo {year} {1976})}\BibitemShut {NoStop}%
\bibitem [{\citenamefont {Bochkov}\ and\ \citenamefont
  {Kuzovlev}(1977)}]{Bochkov1977}%
  \BibitemOpen
  \bibfield  {author} {\bibinfo {author} {\bibfnamefont {G.}~\bibnamefont
  {Bochkov}}\ and\ \bibinfo {author} {\bibfnamefont {Y.~E.}\ \bibnamefont
  {Kuzovlev}},\ }\href@noop {} {\bibfield  {journal} {\bibinfo  {journal} {Zh.
  Eksp. Teor. Fiz}\ }\textbf {\bibinfo {volume} {72}},\ \bibinfo {pages} {238}
  (\bibinfo {year} {1977})}\BibitemShut {NoStop}%
\bibitem [{\citenamefont {Evans}\ \emph {et~al.}(1993)\citenamefont {Evans},
  \citenamefont {Cohen},\ and\ \citenamefont {Morriss}}]{Evans1993}%
  \BibitemOpen
  \bibfield  {author} {\bibinfo {author} {\bibfnamefont {D.~J.}\ \bibnamefont
  {Evans}}, \bibinfo {author} {\bibfnamefont {E.~G.~D.}\ \bibnamefont {Cohen}},
  \ and\ \bibinfo {author} {\bibfnamefont {G.~P.}\ \bibnamefont {Morriss}},\
  }\href {\doibase 10.1103/PhysRevLett.71.2401} {\bibfield  {journal} {\bibinfo
   {journal} {Phys. Rev. Lett.}\ }\textbf {\bibinfo {volume} {71}},\ \bibinfo
  {pages} {2401} (\bibinfo {year} {1993})}\BibitemShut {NoStop}%
\bibitem [{\citenamefont {Gallavotti}\ and\ \citenamefont
  {Cohen}(1995)}]{Gallavotti1995}%
  \BibitemOpen
  \bibfield  {author} {\bibinfo {author} {\bibfnamefont {G.}~\bibnamefont
  {Gallavotti}}\ and\ \bibinfo {author} {\bibfnamefont {E.~G.~D.}\ \bibnamefont
  {Cohen}},\ }\href {\doibase 10.1103/PhysRevLett.74.2694} {\bibfield
  {journal} {\bibinfo  {journal} {Phys. Rev. Lett.}\ }\textbf {\bibinfo
  {volume} {74}},\ \bibinfo {pages} {2694} (\bibinfo {year}
  {1995})}\BibitemShut {NoStop}%
\bibitem [{\citenamefont {Jarzynski}(1997)}]{Jarzynski1997}%
  \BibitemOpen
  \bibfield  {author} {\bibinfo {author} {\bibfnamefont {C.}~\bibnamefont
  {Jarzynski}},\ }\href {\doibase 10.1103/PhysRevLett.78.2690} {\bibfield
  {journal} {\bibinfo  {journal} {Phys. Rev. Lett.}\ }\textbf {\bibinfo
  {volume} {78}},\ \bibinfo {pages} {2690} (\bibinfo {year}
  {1997})}\BibitemShut {NoStop}%
\bibitem [{\citenamefont {Kurchan}(1998)}]{Kurchan1998}%
  \BibitemOpen
  \bibfield  {author} {\bibinfo {author} {\bibfnamefont {J.}~\bibnamefont
  {Kurchan}},\ }\href {\doibase 10.1088/0305-4470/31/16/003} {\bibfield
  {journal} {\bibinfo  {journal} {J. Phys. A: Math. Gen.}\ }\textbf {\bibinfo
  {volume} {31}},\ \bibinfo {pages} {3719} (\bibinfo {year}
  {1998})}\BibitemShut {NoStop}%
\bibitem [{\citenamefont {Lebowitz}\ and\ \citenamefont
  {Spohn}(1999)}]{Lebowitz1999}%
  \BibitemOpen
  \bibfield  {author} {\bibinfo {author} {\bibfnamefont {J.~L.}\ \bibnamefont
  {Lebowitz}}\ and\ \bibinfo {author} {\bibfnamefont {H.}~\bibnamefont
  {Spohn}},\ }\href {\doibase 10.1023/a:1004589714161} {\bibfield  {journal}
  {\bibinfo  {journal} {J. Stat. Phys.}\ }\textbf {\bibinfo {volume} {95}},\
  \bibinfo {pages} {333} (\bibinfo {year} {1999})}\BibitemShut {NoStop}%
\bibitem [{\citenamefont {Maes}(1999)}]{Maes1999}%
  \BibitemOpen
  \bibfield  {author} {\bibinfo {author} {\bibfnamefont {C.}~\bibnamefont
  {Maes}},\ }\href {\doibase 10.1023/a:1004541830999} {\bibfield  {journal}
  {\bibinfo  {journal} {J. Stat. Phys.}\ }\textbf {\bibinfo {volume} {95}},\
  \bibinfo {pages} {367} (\bibinfo {year} {1999})}\BibitemShut {NoStop}%
\bibitem [{\citenamefont {Crooks}(2000)}]{Crooks2000}%
  \BibitemOpen
  \bibfield  {author} {\bibinfo {author} {\bibfnamefont {G.~E.}\ \bibnamefont
  {Crooks}},\ }\href {http://pre.aps.org/abstract/PRE/v61/i3/p2361_1}
  {\bibfield  {journal} {\bibinfo  {journal} {Phys. Rev.E}\ }\textbf {\bibinfo
  {volume} {61}},\ \bibinfo {pages} {2361} (\bibinfo {year}
  {2000})}\BibitemShut {NoStop}%
\bibitem [{\citenamefont {Hatano}\ and\ \citenamefont
  {Sasa}(2001)}]{Hatano2001}%
  \BibitemOpen
  \bibfield  {author} {\bibinfo {author} {\bibfnamefont {T.}~\bibnamefont
  {Hatano}}\ and\ \bibinfo {author} {\bibfnamefont {S.-i.}\ \bibnamefont
  {Sasa}},\ }\href {\doibase 10.1103/PhysRevLett.86.3463} {\bibfield  {journal}
  {\bibinfo  {journal} {Phys. Rev. Lett.}\ }\textbf {\bibinfo {volume} {86}},\
  \bibinfo {pages} {3463} (\bibinfo {year} {2001})}\BibitemShut {NoStop}%
\bibitem [{\citenamefont {Seifert}(2005)}]{Seifert2005}%
  \BibitemOpen
  \bibfield  {author} {\bibinfo {author} {\bibfnamefont {U.}~\bibnamefont
  {Seifert}},\ }\href {\doibase 10.1103/PhysRevLett.95.040602} {\bibfield
  {journal} {\bibinfo  {journal} {Phys. Rev. Lett.}\ }\textbf {\bibinfo
  {volume} {95}},\ \bibinfo {pages} {040602} (\bibinfo {year}
  {2005})}\BibitemShut {NoStop}%
\bibitem [{\citenamefont {Kawai}\ \emph {et~al.}(2007)\citenamefont {Kawai},
  \citenamefont {Parrondo},\ and\ \citenamefont {VandenBroeck}}]{Kawai2007}%
  \BibitemOpen
  \bibfield  {author} {\bibinfo {author} {\bibfnamefont {R.}~\bibnamefont
  {Kawai}}, \bibinfo {author} {\bibfnamefont {J.~M.~R.}\ \bibnamefont
  {Parrondo}}, \ and\ \bibinfo {author} {\bibfnamefont {C.}~\bibnamefont
  {VandenBroeck}},\ }\href {\doibase 10.1103/PhysRevLett.98.080602} {\bibfield
  {journal} {\bibinfo  {journal} {Phys. Rev. Lett.}\ }\textbf {\bibinfo
  {volume} {98}},\ \bibinfo {pages} {080602} (\bibinfo {year}
  {2007})}\BibitemShut {NoStop}%
\bibitem [{\citenamefont {Kurchan}(2000)}]{Kurchan2000}%
  \BibitemOpen
  \bibfield  {author} {\bibinfo {author} {\bibfnamefont {J.}~\bibnamefont
  {Kurchan}},\ }\href {http://arxiv.org/abs/cond-mat/0007360} {\bibfield
  {journal} {\bibinfo  {journal} {arXiv preprint cond-mat/0007360}\ } (\bibinfo
  {year} {2000})}\BibitemShut {NoStop}%
\bibitem [{\citenamefont {Tasaki}(2000)}]{Tasaki2000}%
  \BibitemOpen
  \bibfield  {author} {\bibinfo {author} {\bibfnamefont {H.}~\bibnamefont
  {Tasaki}},\ }\href {http://arxiv.org/abs/cond-mat/0009244} {\bibfield
  {journal} {\bibinfo  {journal} {arXiv preprint cond-mat/0009244}\ } (\bibinfo
  {year} {2000})}\BibitemShut {NoStop}%
\bibitem [{\citenamefont {Yukawa}(2000)}]{Yukawa2000}%
  \BibitemOpen
  \bibfield  {author} {\bibinfo {author} {\bibfnamefont {S.}~\bibnamefont
  {Yukawa}},\ }\href {http://journals.jps.jp/doi/abs/10.1143/JPSJ.69.2367}
  {\bibfield  {journal} {\bibinfo  {journal} {J. Phys. Soc. Jpn.}\ }\textbf
  {\bibinfo {volume} {69}},\ \bibinfo {pages} {2367} (\bibinfo {year}
  {2000})}\BibitemShut {NoStop}%
\bibitem [{\citenamefont {Allahverdyan}\ and\ \citenamefont
  {Nieuwenhuizen}(2005)}]{Allahverdyan2005}%
  \BibitemOpen
  \bibfield  {author} {\bibinfo {author} {\bibfnamefont {A.~E.}\ \bibnamefont
  {Allahverdyan}}\ and\ \bibinfo {author} {\bibfnamefont {T.~M.}\ \bibnamefont
  {Nieuwenhuizen}},\ }\href
  {http://journals.aps.org/pre/abstract/10.1103/PhysRevE.71.066102} {\bibfield
  {journal} {\bibinfo  {journal} {Phys. Rev. E}\ }\textbf {\bibinfo {volume}
  {71}},\ \bibinfo {pages} {066102} (\bibinfo {year} {2005})}\BibitemShut
  {NoStop}%
\bibitem [{\citenamefont {Talkner}\ \emph {et~al.}(2007)\citenamefont
  {Talkner}, \citenamefont {Lutz},\ and\ \citenamefont
  {H\"anggi}}]{Talkner2007}%
  \BibitemOpen
  \bibfield  {author} {\bibinfo {author} {\bibfnamefont {P.}~\bibnamefont
  {Talkner}}, \bibinfo {author} {\bibfnamefont {E.}~\bibnamefont {Lutz}}, \
  and\ \bibinfo {author} {\bibfnamefont {P.}~\bibnamefont {H\"anggi}},\ }\href
  {\doibase 10.1103/PhysRevE.75.050102} {\bibfield  {journal} {\bibinfo
  {journal} {Phys. Rev. E}\ }\textbf {\bibinfo {volume} {75}},\ \bibinfo
  {pages} {050102} (\bibinfo {year} {2007})}\BibitemShut {NoStop}%
\bibitem [{\citenamefont {Andrieux}\ and\ \citenamefont
  {Gaspard}(2008)}]{Andrieux2008}%
  \BibitemOpen
  \bibfield  {author} {\bibinfo {author} {\bibfnamefont {D.}~\bibnamefont
  {Andrieux}}\ and\ \bibinfo {author} {\bibfnamefont {P.}~\bibnamefont
  {Gaspard}},\ }\href {\doibase 10.1103/PhysRevLett.100.230404} {\bibfield
  {journal} {\bibinfo  {journal} {Phys. Rev. Lett.}\ }\textbf {\bibinfo
  {volume} {100}},\ \bibinfo {pages} {230404} (\bibinfo {year}
  {2008})}\BibitemShut {NoStop}%
\bibitem [{\citenamefont {Campisi}\ \emph
  {et~al.}(2011{\natexlab{a}})\citenamefont {Campisi}, \citenamefont
  {H{\"a}nggi},\ and\ \citenamefont {Talkner}}]{Campisi2011}%
  \BibitemOpen
  \bibfield  {author} {\bibinfo {author} {\bibfnamefont {M.}~\bibnamefont
  {Campisi}}, \bibinfo {author} {\bibfnamefont {P.}~\bibnamefont {H{\"a}nggi}},
  \ and\ \bibinfo {author} {\bibfnamefont {P.}~\bibnamefont {Talkner}},\ }\href
  {http://rmp.aps.org/abstract/RMP/v83/i3/p771_1} {\bibfield  {journal}
  {\bibinfo  {journal} {Rev. Mod. Phys.}\ }\textbf {\bibinfo {volume} {83}},\
  \bibinfo {pages} {771} (\bibinfo {year} {2011}{\natexlab{a}})}\BibitemShut
  {NoStop}%
\bibitem [{\citenamefont {Batalh{\~{a}}o}\ \emph {et~al.}(2014)\citenamefont
  {Batalh{\~{a}}o}, \citenamefont {Souza}, \citenamefont {Mazzola},
  \citenamefont {Auccaise}, \citenamefont {Sarthour}, \citenamefont {Oliveira},
  \citenamefont {Goold}, \citenamefont {{De Chiara}}, \citenamefont
  {Paternostro},\ and\ \citenamefont {Serra}}]{Batalhao2014}%
  \BibitemOpen
  \bibfield  {author} {\bibinfo {author} {\bibfnamefont {T.~B.}\ \bibnamefont
  {Batalh{\~{a}}o}}, \bibinfo {author} {\bibfnamefont {A.~M.}\ \bibnamefont
  {Souza}}, \bibinfo {author} {\bibfnamefont {L.}~\bibnamefont {Mazzola}},
  \bibinfo {author} {\bibfnamefont {R.}~\bibnamefont {Auccaise}}, \bibinfo
  {author} {\bibfnamefont {R.~S.}\ \bibnamefont {Sarthour}}, \bibinfo {author}
  {\bibfnamefont {I.~S.}\ \bibnamefont {Oliveira}}, \bibinfo {author}
  {\bibfnamefont {J.}~\bibnamefont {Goold}}, \bibinfo {author} {\bibfnamefont
  {G.}~\bibnamefont {{De Chiara}}}, \bibinfo {author} {\bibfnamefont
  {M.}~\bibnamefont {Paternostro}}, \ and\ \bibinfo {author} {\bibfnamefont
  {R.~M.}\ \bibnamefont {Serra}},\ }\href@noop {} {\bibfield  {journal}
  {\bibinfo  {journal} {Phys. Rev. Lett.}\ }\textbf {\bibinfo {volume} {113}},\
  \bibinfo {pages} {140601} (\bibinfo {year} {2014})}\BibitemShut {NoStop}%
\bibitem [{\citenamefont {An}\ \emph {et~al.}(2015)\citenamefont {An},
  \citenamefont {Zhang}, \citenamefont {Um}, \citenamefont {Lv}, \citenamefont
  {Lu}, \citenamefont {Zhang}, \citenamefont {Yi}, \citenamefont {Quan},\ and\
  \citenamefont {Kim}}]{ShuomingAn2015}%
  \BibitemOpen
  \bibfield  {author} {\bibinfo {author} {\bibfnamefont {S.-M.}\ \bibnamefont
  {An}}, \bibinfo {author} {\bibfnamefont {J.-N.}\ \bibnamefont {Zhang}},
  \bibinfo {author} {\bibfnamefont {M.}~\bibnamefont {Um}}, \bibinfo {author}
  {\bibfnamefont {D.-S.}\ \bibnamefont {Lv}}, \bibinfo {author} {\bibfnamefont
  {Y.}~\bibnamefont {Lu}}, \bibinfo {author} {\bibfnamefont {J.-H.}\
  \bibnamefont {Zhang}}, \bibinfo {author} {\bibfnamefont {Z.-Q.}\ \bibnamefont
  {Yi}}, \bibinfo {author} {\bibfnamefont {H.-T.}\ \bibnamefont {Quan}}, \ and\
  \bibinfo {author} {\bibfnamefont {K.}~\bibnamefont {Kim}},\ }\href@noop {}
  {\bibfield  {journal} {\bibinfo  {journal} {Nat. Phys.}\ }\textbf {\bibinfo
  {volume} {11}},\ \bibinfo {pages} {193} (\bibinfo {year} {2015})}\BibitemShut
  {NoStop}%
\bibitem [{\citenamefont {Jarzynski}\ \emph {et~al.}(2015)\citenamefont
  {Jarzynski}, \citenamefont {Quan},\ and\ \citenamefont
  {Rahav}}]{Jarzynski2015}%
  \BibitemOpen
  \bibfield  {author} {\bibinfo {author} {\bibfnamefont {C.}~\bibnamefont
  {Jarzynski}}, \bibinfo {author} {\bibfnamefont {H.~T.}~\bibnamefont {Quan}}, \
  and\ \bibinfo {author} {\bibfnamefont {S.}~\bibnamefont {Rahav}},\
  }\href@noop {} {\bibfield  {journal} {\bibinfo  {journal} {Phys. Rev. X}\
  }\textbf {\bibinfo {volume} {5}},\ \bibinfo {pages} {031038} (\bibinfo {year}
  {2015})}\BibitemShut {NoStop}%
\bibitem [{\citenamefont {Boixo}\ \emph {et~al.}(2014)\citenamefont {Boixo},
  \citenamefont {R{\o}nnow}, \citenamefont {Isakov}, \citenamefont {Wang},
  \citenamefont {Wecker}, \citenamefont {Lidar}, \citenamefont {Martinis},\
  and\ \citenamefont {Troyer}}]{Boixo2014}%
  \BibitemOpen
  \bibfield  {author} {\bibinfo {author} {\bibfnamefont {S.}~\bibnamefont
  {Boixo}}, \bibinfo {author} {\bibfnamefont {T.~F.}\ \bibnamefont
  {R{\o}nnow}}, \bibinfo {author} {\bibfnamefont {S.~V.}\ \bibnamefont
  {Isakov}}, \bibinfo {author} {\bibfnamefont {Z.}~\bibnamefont {Wang}},
  \bibinfo {author} {\bibfnamefont {D.}~\bibnamefont {Wecker}}, \bibinfo
  {author} {\bibfnamefont {D.~A.}\ \bibnamefont {Lidar}}, \bibinfo {author}
  {\bibfnamefont {J.~M.}\ \bibnamefont {Martinis}}, \ and\ \bibinfo {author}
  {\bibfnamefont {M.}~\bibnamefont {Troyer}},\ }\href {\doibase
  10.1038/nphys2900} {\bibfield  {journal} {\bibinfo  {journal} {Nat. Phys.}\
  }\textbf {\bibinfo {volume} {10}},\ \bibinfo {pages} {218} (\bibinfo {year}
  {2014})}\BibitemShut {NoStop}%
\bibitem [{\citenamefont {Kosloff}(2013)}]{Kosloff2013}%
  \BibitemOpen
  \bibfield  {author} {\bibinfo {author} {\bibfnamefont {R.}~\bibnamefont
  {Kosloff}},\ }\href@noop {} {\bibfield  {journal} {\bibinfo  {journal}
  {Entropy}\ }\textbf {\bibinfo {volume} {15}},\ \bibinfo {pages} {2100}
  (\bibinfo {year} {2013})}\BibitemShut {NoStop}%
\bibitem [{\citenamefont {Gemmer}\ \emph {et~al.}(2005)\citenamefont {Gemmer},
  \citenamefont {Michel},\ and\ \citenamefont {Mahler}}]{Gemmer2005}%
  \BibitemOpen
  \bibfield  {author} {\bibinfo {author} {\bibfnamefont {J.}~\bibnamefont
  {Gemmer}}, \bibinfo {author} {\bibfnamefont {M.}~\bibnamefont {Michel}}, \
  and\ \bibinfo {author} {\bibfnamefont {G.}~\bibnamefont {Mahler}},\
  }\href@noop {} {\bibfield  {journal} {\bibinfo  {journal} {Lecture Notes in
  Physics}\ }\textbf {\bibinfo {volume} {29}},\ \bibinfo {pages} {53} (\bibinfo
  {year} {2005})}\BibitemShut {NoStop}%
\bibitem [{\citenamefont {Breuer}\ and\ \citenamefont
  {Petruccione}(2002)}]{Breuer2002}%
  \BibitemOpen
  \bibfield  {author} {\bibinfo {author} {\bibfnamefont {H.-P.}\ \bibnamefont
  {Breuer}}\ and\ \bibinfo {author} {\bibfnamefont {F.}~\bibnamefont
  {Petruccione}},\ }\href@noop {} {\emph {\bibinfo {title} {The theory of open
  quantum systems}}}\ (\bibinfo  {publisher} {Oxford university press},\
  \bibinfo {year} {2002})\BibitemShut {NoStop}%
\bibitem [{\citenamefont {Alicki}\ and\ \citenamefont
  {Lendi}(2010)}]{Alicki2010}%
  \BibitemOpen
  \bibfield  {author} {\bibinfo {author} {\bibfnamefont {B.~R.}\ \bibnamefont
  {Alicki}}\ and\ \bibinfo {author} {\bibfnamefont {K.}~\bibnamefont {Lendi}},\
  }\href@noop {} {\emph {\bibinfo {title} {Quantum Dynamical Semigroups and
  Applications}}}\ (\bibinfo  {publisher} {Springer, Berlin},\ \bibinfo {year}
  {2010})\BibitemShut {NoStop}%
\bibitem [{\citenamefont {Rivas}\ and\ \citenamefont
  {Huelga}(2012)}]{Rivas2012}%
  \BibitemOpen
  \bibfield  {author} {\bibinfo {author} {\bibfnamefont {A.}~\bibnamefont
  {Rivas}}\ and\ \bibinfo {author} {\bibfnamefont {S.~F.}\ \bibnamefont
  {Huelga}},\ }\href {\doibase 10.1007/978-3-642-23354-8} {\emph {\bibinfo
  {title} {Open Quantum Systems}}}\ (\bibinfo  {publisher} {Springer Berlin
  Heidelberg},\ \bibinfo {year} {2012})\BibitemShut {NoStop}%
\bibitem [{\citenamefont {Spohn}\ and\ \citenamefont
  {Lebowitz}(1978)}]{Spohn1978}%
  \BibitemOpen
  \bibfield  {author} {\bibinfo {author} {\bibfnamefont {H.}~\bibnamefont
  {Spohn}}\ and\ \bibinfo {author} {\bibfnamefont {J.~L.}\ \bibnamefont
  {Lebowitz}},\ }in\ \href@noop {} {\emph {\bibinfo {booktitle} {Adv. Chem.
  Phys}}}\ (\bibinfo {year} {1978})\BibitemShut {NoStop}%
\bibitem [{\citenamefont {Davies}\ and\ \citenamefont
  {Spohn}(1978)}]{Davies1978}%
  \BibitemOpen
  \bibfield  {author} {\bibinfo {author} {\bibfnamefont {E.~B.}\ \bibnamefont
  {Davies}}\ and\ \bibinfo {author} {\bibfnamefont {H.}~\bibnamefont {Spohn}},\
  }\href@noop {} {\bibfield  {journal} {\bibinfo  {journal} {J. Stat. Phys.}\
  }\textbf {\bibinfo {volume} {19}},\ \bibinfo {pages} {511} (\bibinfo {year}
  {1978})}\BibitemShut {NoStop}%
\bibitem [{\citenamefont {Alicki}(1979)}]{Alicki1979}%
  \BibitemOpen
  \bibfield  {author} {\bibinfo {author} {\bibfnamefont {R.}~\bibnamefont
  {Alicki}},\ }\href {http://iopscience.iop.org/0305-4470/12/5/007} {\bibfield
  {journal} {\bibinfo  {journal} {J. Phys. A: Math. Theor.}\ }\textbf {\bibinfo
  {volume} {12}},\ \bibinfo {pages} {L103} (\bibinfo {year}
  {1979})}\BibitemShut {NoStop}%
\bibitem [{\citenamefont {Gardiner}(1983)}]{Gardiner1983}%
  \BibitemOpen
  \bibfield  {author} {\bibinfo {author} {\bibfnamefont {C.~W.}\ \bibnamefont
  {Gardiner}},\ }\href {\doibase 10.1007/978-3-662-02377-8} {\emph {\bibinfo
  {title} {Handbook of Stochastic Methods}}}\ (\bibinfo  {publisher} {Springer
  Berlin Heidelberg},\ \bibinfo {year} {1983})\BibitemShut {NoStop}%
\bibitem [{\citenamefont {Risken}(1984)}]{Risken1984}%
  \BibitemOpen
  \bibfield  {author} {\bibinfo {author} {\bibfnamefont {H.~H.}\ \bibnamefont
  {Risken}},\ }\href {http://opac.inria.fr/record=b1092006} {\emph {\bibinfo
  {title} {The Fokker-Planck equation : methods of solution and
  applications}}}\ (\bibinfo  {publisher} {Springer-Verlag},\ \bibinfo
  {address} {Berlin, New York},\ \bibinfo {year} {1984})\BibitemShut {NoStop}%
\bibitem [{Note1()}]{Note1}%
  \BibitemOpen
  \bibinfo {note} {In some studies, e.g., Refs.~\cite {Esposito2009},~\cite
  {Garrahan2010},~\cite {Gasparinetti2014},~\cite {Cuetara2015}, and~\cite
  {Pigeon2015}, the moment-generating function rather than the CF was used. If
  all of the moments of a stochastic quantity exist and are finite, which we
  assume throughout this article, there are no essential differences between
  two functions. A moment-generating function may be simply regarded as a CF
  evaluated on the imaginary axis~\cite {Gardiner1983}.}\BibitemShut {Stop}%
\bibitem [{\citenamefont {Imparato}\ and\ \citenamefont
  {Peliti}(2007)}]{Imparato2007}%
  \BibitemOpen
  \bibfield  {author} {\bibinfo {author} {\bibfnamefont {A.}~\bibnamefont
  {Imparato}}\ and\ \bibinfo {author} {\bibfnamefont {L.}~\bibnamefont
  {Peliti}},\ }\href {\doibase 10.1016/j.crhy.2007.04.017} {\bibfield
  {journal} {\bibinfo  {journal} {Comptes Rendus Physique}\ }\textbf {\bibinfo
  {volume} {8}},\ \bibinfo {pages} {556} (\bibinfo {year} {2007})}\BibitemShut
  {NoStop}%
\bibitem [{\citenamefont {Touchette}(2008)}]{Touchette2008}%
  \BibitemOpen
  \bibfield  {author} {\bibinfo {author} {\bibfnamefont {H.}~\bibnamefont
  {Touchette}},\ }\href@noop {} {\bibfield  {journal} {\bibinfo  {journal}
  {Phys. Rep.}\ }\textbf {\bibinfo {volume} {478}},\ \bibinfo {pages} {1}
  (\bibinfo {year} {2008})}\BibitemShut {NoStop}%
\bibitem [{\citenamefont {Garrahan}\ and\ \citenamefont
  {Lesanovsky}(2010)}]{Garrahan2010}%
  \BibitemOpen
  \bibfield  {author} {\bibinfo {author} {\bibfnamefont {J.~P.}\ \bibnamefont
  {Garrahan}}\ and\ \bibinfo {author} {\bibfnamefont {I.}~\bibnamefont
  {Lesanovsky}},\ }\href {\doibase 10.1103/PhysRevLett.104.160601} {\bibfield
  {journal} {\bibinfo  {journal} {Phys. Rev. Lett.}\ }\textbf {\bibinfo
  {volume} {104}},\ \bibinfo {pages} {160601} (\bibinfo {year}
  {2010})}\BibitemShut {NoStop}%
\bibitem [{\citenamefont {\ifmmode \check{Z}\else
  \v{Z}\fi{}nidari\ifmmode~\check{c}\else
  \v{c}\fi{}}(2014)}]{Zinidarifmmodeheckclseci2014}%
  \BibitemOpen
  \bibfield  {author} {\bibinfo {author} {\bibfnamefont {M.}~\bibnamefont
  {\ifmmode \check{Z}\else \v{Z}\fi{}nidari\ifmmode~\check{c}\else
  \v{c}\fi{}}},\ }\href {\doibase 10.1103/PhysRevLett.112.040602} {\bibfield
  {journal} {\bibinfo  {journal} {Phys. Rev. Lett.}\ }\textbf {\bibinfo
  {volume} {112}},\ \bibinfo {pages} {040602} (\bibinfo {year}
  {2014})}\BibitemShut {NoStop}%
\bibitem [{\citenamefont {Pigeon}\ \emph {et~al.}(2015)\citenamefont {Pigeon},
  \citenamefont {Fusco}, \citenamefont {Xuereb}, \citenamefont {Chiara},\ and\
  \citenamefont {Paternostro}}]{Pigeon2015}%
  \BibitemOpen
  \bibfield  {author} {\bibinfo {author} {\bibfnamefont {S.}~\bibnamefont
  {Pigeon}}, \bibinfo {author} {\bibfnamefont {L.}~\bibnamefont {Fusco}},
  \bibinfo {author} {\bibfnamefont {A.}~\bibnamefont {Xuereb}}, \bibinfo
  {author} {\bibfnamefont {G.~D.}\ \bibnamefont {Chiara}}, \ and\ \bibinfo
  {author} {\bibfnamefont {M.}~\bibnamefont {Paternostro}},\ }\href {\doibase
  10.1088/1367-2630/18/1/013009} {\bibfield  {journal} {\bibinfo  {journal}
  {New J. Phys.}\ }\textbf {\bibinfo {volume} {18}},\ \bibinfo {pages} {013009}
  (\bibinfo {year} {2015})}\BibitemShut {NoStop}%
\bibitem [{\citenamefont {Seifert}(2011)}]{Seifert2011}%
  \BibitemOpen
  \bibfield  {author} {\bibinfo {author} {\bibfnamefont {U.}~\bibnamefont
  {Seifert}},\ }in\ \href@noop {} {\emph {\bibinfo {booktitle} {Nonequilibrium
  Statistical Physics Today: Granada Seminar on Computational \& Statistical
  Physics}}}\ (\bibinfo {year} {2011})\ pp.\ \bibinfo {pages}
  {56--76}\BibitemShut {NoStop}%
\bibitem [{\citenamefont {Rousochatzakis}\ and\ \citenamefont
  {Luban}(2005)}]{Rousochatzakis2005}%
  \BibitemOpen
  \bibfield  {author} {\bibinfo {author} {\bibfnamefont {I.}~\bibnamefont
  {Rousochatzakis}}\ and\ \bibinfo {author} {\bibfnamefont {M.}~\bibnamefont
  {Luban}},\ }\href {http://prb.aps.org/abstract/PRB/v72/i13/e134424}
  {\bibfield  {journal} {\bibinfo  {journal} {Phys. Rev. B}\ }\textbf {\bibinfo
  {volume} {72}},\ \bibinfo {pages} {134424} (\bibinfo {year}
  {2005})}\BibitemShut {NoStop}%
\bibitem [{\citenamefont {Cai}\ \emph {et~al.}(2010)\citenamefont {Cai},
  \citenamefont {Popescu},\ and\ \citenamefont {Briegel}}]{Cai2010}%
  \BibitemOpen
  \bibfield  {author} {\bibinfo {author} {\bibfnamefont {J.}~\bibnamefont
  {Cai}}, \bibinfo {author} {\bibfnamefont {S.}~\bibnamefont {Popescu}}, \ and\
  \bibinfo {author} {\bibfnamefont {H.~J.}\ \bibnamefont {Briegel}},\ }\href
  {http://pre.aps.org/abstract/PRE/v82/i2/e021921} {\bibfield  {journal}
  {\bibinfo  {journal} {Phys. Rev. E}\ }\textbf {\bibinfo {volume} {82}},\
  \bibinfo {pages} {021921} (\bibinfo {year} {2010})}\BibitemShut {NoStop}%
\bibitem [{\citenamefont {Carmichael}(1993)}]{Carmichael1993}%
  \BibitemOpen
  \bibfield  {author} {\bibinfo {author} {\bibfnamefont {H.}~\bibnamefont
  {Carmichael}},\ }\href@noop {} {\emph {\bibinfo {title} {An open systems
  approach to Quantum Optics: lectures presented at the Universit{\'e} Libre de
  Bruxelles, October 28 to November 4, 1991}}},\ Vol.~\bibinfo {volume} {18}\
  (\bibinfo  {publisher} {Springer},\ \bibinfo {year} {1993})\BibitemShut
  {NoStop}%
\bibitem [{\citenamefont {Plenio}\ and\ \citenamefont
  {Knight}(1998)}]{Plenio1998}%
  \BibitemOpen
  \bibfield  {author} {\bibinfo {author} {\bibfnamefont {M.}~\bibnamefont
  {Plenio}}\ and\ \bibinfo {author} {\bibfnamefont {P.}~\bibnamefont
  {Knight}},\ }\href {http://rmp.aps.org/abstract/RMP/v70/i1/p101_1} {\bibfield
   {journal} {\bibinfo  {journal} {Rev. Mod. Phys.}\ }\textbf {\bibinfo
  {volume} {70}},\ \bibinfo {pages} {101} (\bibinfo {year} {1998})}\BibitemShut
  {NoStop}%
\bibitem [{\citenamefont {Wiseman}\ and\ \citenamefont
  {Milburn}(2010)}]{Wiseman2010}%
  \BibitemOpen
  \bibfield  {author} {\bibinfo {author} {\bibfnamefont {H.~M.}\ \bibnamefont
  {Wiseman}}\ and\ \bibinfo {author} {\bibfnamefont {G.~J.}\ \bibnamefont
  {Milburn}},\ }\href@noop {} {\emph {\bibinfo {title} {Quantum measurement and
  control}}}\ (\bibinfo  {publisher} {Cambridge University Press},\ \bibinfo
  {year} {2010})\BibitemShut {NoStop}%
\bibitem [{\citenamefont {Breuer}(2003)}]{Breuer2003}%
  \BibitemOpen
  \bibfield  {author} {\bibinfo {author} {\bibfnamefont {H.-P.}\ \bibnamefont
  {Breuer}},\ }\href {\doibase 10.1103/PhysRevA.68.032105} {\bibfield
  {journal} {\bibinfo  {journal} {Phys. Rev. A}\ }\textbf {\bibinfo {volume}
  {68}},\ \bibinfo {pages} {032105} (\bibinfo {year} {2003})}\BibitemShut
  {NoStop}%
\bibitem [{\citenamefont {Derezi{\'n}ski}\ \emph {et~al.}(2008)\citenamefont
  {Derezi{\'n}ski}, \citenamefont {De~Roeck},\ and\ \citenamefont
  {Maes}}]{Derezinski2008}%
  \BibitemOpen
  \bibfield  {author} {\bibinfo {author} {\bibfnamefont {J.}~\bibnamefont
  {Derezi{\'n}ski}}, \bibinfo {author} {\bibfnamefont {W.}~\bibnamefont
  {De~Roeck}}, \ and\ \bibinfo {author} {\bibfnamefont {C.}~\bibnamefont
  {Maes}},\ }\href@noop {} {\bibfield  {journal} {\bibinfo  {journal} {J. Stat.
  Phys.}\ }\textbf {\bibinfo {volume} {131}},\ \bibinfo {pages} {341} (\bibinfo
  {year} {2008})}\BibitemShut {NoStop}%
\bibitem [{\citenamefont {Elouard}\ \emph {et~al.}(2015)\citenamefont
  {Elouard}, \citenamefont {Auff¨¨ves},\ and\ \citenamefont
  {Clusel}}]{Elouard2015}%
  \BibitemOpen
  \bibfield  {author} {\bibinfo {author} {\bibfnamefont {C.}~\bibnamefont
  {Elouard}}, \bibinfo {author} {\bibfnamefont {A.}~\bibnamefont {Auff¨¨ves}},
  \ and\ \bibinfo {author} {\bibfnamefont {M.}~\bibnamefont {Clusel}},\
  }\href@noop {} {\  (\bibinfo {year} {2015})},\ \Eprint
  {http://arxiv.org/abs/1507.00312} {1507.00312} \BibitemShut {NoStop}%
\bibitem [{\citenamefont {Liu}(2016)}]{Liu2016}%
  \BibitemOpen
  \bibfield  {author} {\bibinfo {author} {\bibfnamefont {F.}~\bibnamefont
  {Liu}},\ }\href {\doibase 10.1103/PhysRevE.93.012127} {\bibfield  {journal}
  {\bibinfo  {journal} {Phys. Rev. E}\ }\textbf {\bibinfo {volume} {93}},\
  \bibinfo {pages} {012127} (\bibinfo {year} {2016})}\BibitemShut {NoStop}%
\bibitem [{Note2()}]{Note2}%
  \BibitemOpen
  \bibinfo {note} {The closed quantum systems may be regarded as a type of QME.
  Because of trivialness, we do not consider it.}\BibitemShut {Stop}%
\bibitem [{\citenamefont {Lindblad}(1975)}]{Lindblad1975}%
  \BibitemOpen
  \bibfield  {author} {\bibinfo {author} {\bibfnamefont {G.}~\bibnamefont
  {Lindblad}},\ }\href {http://link.springer.com/article/10.1007/BF01609396}
  {\bibfield  {journal} {\bibinfo  {journal} {Comm. Math. Phys.}\ }\textbf
  {\bibinfo {volume} {40}},\ \bibinfo {pages} {147} (\bibinfo {year}
  {1975})}\BibitemShut {NoStop}%
\bibitem [{\citenamefont {Schaller}(2014)}]{Schaller2014}%
  \BibitemOpen
  \bibfield  {author} {\bibinfo {author} {\bibfnamefont {G.}~\bibnamefont
  {Schaller}},\ }\href {\doibase 10.1007/978-3-319-03877-3} {\emph {\bibinfo
  {title} {Open Quantum Systems Far from Equilibrium}}}\ (\bibinfo  {publisher}
  {Springer International Publishing},\ \bibinfo {year} {2014})\BibitemShut
  {NoStop}%
\bibitem [{\citenamefont {Geva}\ and\ \citenamefont
  {Kosloff}(1994)}]{Geva1994}%
  \BibitemOpen
  \bibfield  {author} {\bibinfo {author} {\bibfnamefont {E.}~\bibnamefont
  {Geva}}\ and\ \bibinfo {author} {\bibfnamefont {R.}~\bibnamefont {Kosloff}},\
  }\href {http://journals.aps.org/pre/abstract/10.1103/PhysRevE.49.3903}
  {\bibfield  {journal} {\bibinfo  {journal} {Phys. Rev. E}\ }\textbf {\bibinfo
  {volume} {49}},\ \bibinfo {pages} {3903} (\bibinfo {year}
  {1994})}\BibitemShut {NoStop}%
\bibitem [{\citenamefont {Mollow}(1975)}]{Mollow1975}%
  \BibitemOpen
  \bibfield  {author} {\bibinfo {author} {\bibfnamefont {B.~R.}\ \bibnamefont
  {Mollow}},\ }\href {\doibase 10.1103/PhysRevA.12.1919} {\bibfield  {journal}
  {\bibinfo  {journal} {Phys. Rev. A}\ }\textbf {\bibinfo {volume} {12}},\
  \bibinfo {pages} {1919} (\bibinfo {year} {1975})}\BibitemShut {NoStop}%
\bibitem [{\citenamefont {Albash}\ \emph {et~al.}(2012)\citenamefont {Albash},
  \citenamefont {Boixo}, \citenamefont {Lidar},\ and\ \citenamefont
  {Zanardi}}]{Albash2012}%
  \BibitemOpen
  \bibfield  {author} {\bibinfo {author} {\bibfnamefont {T.}~\bibnamefont
  {Albash}}, \bibinfo {author} {\bibfnamefont {S.}~\bibnamefont {Boixo}},
  \bibinfo {author} {\bibfnamefont {D.~A.}\ \bibnamefont {Lidar}}, \ and\
  \bibinfo {author} {\bibfnamefont {P.}~\bibnamefont {Zanardi}},\ }\href
  {http://iopscience.iop.org/1367-2630/14/12/123016} {\bibfield  {journal}
  {\bibinfo  {journal} {New J. Phys.}\ }\textbf {\bibinfo {volume} {14}},\
  \bibinfo {pages} {123016} (\bibinfo {year} {2012})}\BibitemShut {NoStop}%
\bibitem [{\citenamefont {Bl{\"{u}}mel}\ \emph {et~al.}(1991)\citenamefont
  {Bl{\"{u}}mel}, \citenamefont {A.Buchleitner}, \citenamefont {Graham},
  \citenamefont {Sirko}, \citenamefont {Smilansky},\ and\ \citenamefont
  {Walther}}]{Bluemel1991}%
  \BibitemOpen
  \bibfield  {author} {\bibinfo {author} {\bibfnamefont {R.}~\bibnamefont
  {Bl{\"{u}}mel}}, \bibinfo {author} {\bibnamefont {A.Buchleitner}}, \bibinfo
  {author} {\bibfnamefont {R.}~\bibnamefont {Graham}}, \bibinfo {author}
  {\bibfnamefont {L.}~\bibnamefont {Sirko}}, \bibinfo {author} {\bibfnamefont
  {U.}~\bibnamefont {Smilansky}}, \ and\ \bibinfo {author} {\bibfnamefont
  {H.}~\bibnamefont {Walther}},\ }\href@noop {} {\bibfield  {journal} {\bibinfo
   {journal} {Phys Rev A}\ }\textbf {\bibinfo {volume} {44}},\ \bibinfo {pages}
  {4521} (\bibinfo {year} {1991})}\BibitemShut {NoStop}%
\bibitem [{\citenamefont {Kohler}\ \emph {et~al.}(1997)\citenamefont {Kohler},
  \citenamefont {Dittrich},\ and\ \citenamefont {H{\"a}nggi}}]{Kohler1997}%
  \BibitemOpen
  \bibfield  {author} {\bibinfo {author} {\bibfnamefont {S.}~\bibnamefont
  {Kohler}}, \bibinfo {author} {\bibfnamefont {T.}~\bibnamefont {Dittrich}}, \
  and\ \bibinfo {author} {\bibfnamefont {P.}~\bibnamefont {H{\"a}nggi}},\
  }\href {http://journals.aps.org/pre/abstract/10.1103/PhysRevE.55.300}
  {\bibfield  {journal} {\bibinfo  {journal} {Phys. Rev. E}\ }\textbf {\bibinfo
  {volume} {55}},\ \bibinfo {pages} {300} (\bibinfo {year} {1997})}\BibitemShut
  {NoStop}%
\bibitem [{\citenamefont {Breuer}\ and\ \citenamefont
  {Petruccione}(1997)}]{Breuer1997}%
  \BibitemOpen
  \bibfield  {author} {\bibinfo {author} {\bibfnamefont {H.-P.}\ \bibnamefont
  {Breuer}}\ and\ \bibinfo {author} {\bibfnamefont {F.}~\bibnamefont
  {Petruccione}},\ }\href
  {http://journals.aps.org/pra/abstract/10.1103/PhysRevA.55.3101} {\bibfield
  {journal} {\bibinfo  {journal} {Phys. Rev. A}\ }\textbf {\bibinfo {volume}
  {55}},\ \bibinfo {pages} {3101} (\bibinfo {year} {1997})}\BibitemShut
  {NoStop}%
\bibitem [{\citenamefont {Szczygielski}\ \emph {et~al.}(2013)\citenamefont
  {Szczygielski}, \citenamefont {Gelbwaser-Klimovsky},\ and\ \citenamefont
  {Alicki}}]{Szczygielski2013}%
  \BibitemOpen
  \bibfield  {author} {\bibinfo {author} {\bibfnamefont {K.}~\bibnamefont
  {Szczygielski}}, \bibinfo {author} {\bibfnamefont {D.}~\bibnamefont
  {Gelbwaser-Klimovsky}}, \ and\ \bibinfo {author} {\bibfnamefont
  {R.}~\bibnamefont {Alicki}},\ }\href
  {http://journals.aps.org/pre/abstract/10.1103/PhysRevE.87.012120} {\bibfield
  {journal} {\bibinfo  {journal} {Phys. Rev. E}\ }\textbf {\bibinfo {volume}
  {87}},\ \bibinfo {pages} {012120} (\bibinfo {year} {2013})}\BibitemShut
  {NoStop}%
\bibitem [{\citenamefont {Alicki}\ \emph {et~al.}(2006)\citenamefont {Alicki},
  \citenamefont {Lidar},\ and\ \citenamefont {Zanardi}}]{Alicki2006}%
  \BibitemOpen
  \bibfield  {author} {\bibinfo {author} {\bibfnamefont {R.}~\bibnamefont
  {Alicki}}, \bibinfo {author} {\bibfnamefont {D.~A.}\ \bibnamefont {Lidar}}, \
  and\ \bibinfo {author} {\bibfnamefont {P.}~\bibnamefont {Zanardi}},\ }\href
  {http://pra.aps.org/abstract/PRA/v73/i5/e052311} {\bibfield  {journal}
  {\bibinfo  {journal} {Phys. Rev. A}\ }\textbf {\bibinfo {volume} {73}},\
  \bibinfo {pages} {052311} (\bibinfo {year} {2006})}\BibitemShut {NoStop}%
\bibitem [{\citenamefont {Sambe}(1973)}]{Sambe1973}%
  \BibitemOpen
  \bibfield  {author} {\bibinfo {author} {\bibfnamefont {H.}~\bibnamefont
  {Sambe}},\ }\href@noop {} {\bibfield  {journal} {\bibinfo  {journal} {Phys.
  Rev. A}\ }\textbf {\bibinfo {volume} {7}},\ \bibinfo {pages} {2203} (\bibinfo
  {year} {1973})}\BibitemShut {NoStop}%
\bibitem [{\citenamefont {Shirley}(1965)}]{Shirley1965}%
  \BibitemOpen
  \bibfield  {author} {\bibinfo {author} {\bibfnamefont {J.~H.}\ \bibnamefont
  {Shirley}},\ }\href {\doibase 10.1103/PhysRev.138.B979} {\bibfield  {journal}
  {\bibinfo  {journal} {Phys. Rev.}\ }\textbf {\bibinfo {volume} {138}},\
  \bibinfo {pages} {B979} (\bibinfo {year} {1965})}\BibitemShut {NoStop}%
\bibitem [{\citenamefont {Zeldovich}(1967)}]{Zeldovich1967}%
  \BibitemOpen
  \bibfield  {author} {\bibinfo {author} {\bibfnamefont {Y.~B.}\ \bibnamefont
  {Zeldovich}},\ }\href@noop {} {\bibfield  {journal} {\bibinfo  {journal}
  {Sov. Phys. JETP}\ }\textbf {\bibinfo {volume} {24}},\ \bibinfo {pages}
  {1006} (\bibinfo {year} {1967})}\BibitemShut {NoStop}%
\bibitem [{\citenamefont {Wiseman}\ and\ \citenamefont
  {Milburn}(1993)}]{Wiseman1993}%
  \BibitemOpen
  \bibfield  {author} {\bibinfo {author} {\bibfnamefont {H.~M.}~\bibnamefont
  {Wiseman}}\ and\ \bibinfo {author} {\bibfnamefont {G.~J.}~\bibnamefont
  {Milburn}},\ }\href@noop {} {\bibfield  {journal} {\bibinfo  {journal} {Phys.
  Rev. A}\ }\textbf {\bibinfo {volume} {47}},\ \bibinfo {pages} {1652}
  (\bibinfo {year} {1993})}\BibitemShut {NoStop}%
\bibitem [{\citenamefont {Kist}\ \emph {et~al.}(1999)\citenamefont {Kist},
  \citenamefont {Orszag}, \citenamefont {Brun},\ and\ \citenamefont
  {Davidovich}}]{Kist1999}%
  \BibitemOpen
  \bibfield  {author} {\bibinfo {author} {\bibfnamefont {T.~B.~L.}\
  \bibnamefont {Kist}}, \bibinfo {author} {\bibfnamefont {M.}~\bibnamefont
  {Orszag}}, \bibinfo {author} {\bibfnamefont {T.~A.}\ \bibnamefont {Brun}}, \
  and\ \bibinfo {author} {\bibfnamefont {L.}~\bibnamefont {Davidovich}},\
  }\href {http://stacks.iop.org/1464-4266/1/i=2/a=009} {\bibfield  {journal}
  {\bibinfo  {journal} {J. Opt. B: Quantum Semiclass. Opt.}\ }\textbf {\bibinfo
  {volume} {1}},\ \bibinfo {pages} {251} (\bibinfo {year} {1999})}\BibitemShut
  {NoStop}%
\bibitem [{Note3()}]{Note3}%
  \BibitemOpen
  \bibinfo {note} {Note that the separation of the QME~(\ref {masterequation})
  is not unique. Different separations may correspond different experimental
  monitoring schemes of the open system~\cite {Kist1999}.}\BibitemShut {Stop}%
\bibitem [{\citenamefont {Campisi}\ \emph
  {et~al.}(2011{\natexlab{b}})\citenamefont {Campisi}, \citenamefont
  {Talkner},\ and\ \citenamefont {H{\"a}nggi}}]{Campisi2011a}%
  \BibitemOpen
  \bibfield  {author} {\bibinfo {author} {\bibfnamefont {M.}~\bibnamefont
  {Campisi}}, \bibinfo {author} {\bibfnamefont {P.}~\bibnamefont {Talkner}}, \
  and\ \bibinfo {author} {\bibfnamefont {P.}~\bibnamefont {H{\"a}nggi}},\
  }\href {http://rsta.royalsocietypublishing.org/content/369/1935/291.short}
  {\bibfield  {journal} {\bibinfo  {journal} {Phil. Trans. R. Soc. A}\ }\textbf
  {\bibinfo {volume} {369}},\ \bibinfo {pages} {291} (\bibinfo {year}
  {2011}{\natexlab{b}})}\BibitemShut {NoStop}%
\bibitem [{\citenamefont {Liu}\ and\ \citenamefont {Ouyang}(2014)}]{Liu2014b}%
  \BibitemOpen
  \bibfield  {author} {\bibinfo {author} {\bibfnamefont {F.}~\bibnamefont
  {Liu}}\ and\ \bibinfo {author} {\bibfnamefont {Z.-C.}\ \bibnamefont
  {Ouyang}},\ }\href@noop {} {\bibfield  {journal} {\bibinfo  {journal} {Chin.
  Phys. B}\ }\textbf {\bibinfo {volume} {23}},\ \bibinfo {pages} {070512}
  (\bibinfo {year} {2014})}\BibitemShut {NoStop}%
\bibitem [{\citenamefont {Jarzynski}(2007)}]{Jarzynski2007}%
  \BibitemOpen
  \bibfield  {author} {\bibinfo {author} {\bibfnamefont {C.}~\bibnamefont
  {Jarzynski}},\ }\href
  {http://www.sciencedirect.com/science/article/pii/S1631070507000734}
  {\bibfield  {journal} {\bibinfo  {journal} {C. R. Phys.}\ }\textbf {\bibinfo
  {volume} {8}},\ \bibinfo {pages} {495} (\bibinfo {year} {2007})}\BibitemShut
  {NoStop}%
\bibitem [{\citenamefont {Sakurai}(1994)}]{Sakurai1994}%
  \BibitemOpen
  \bibfield  {author} {\bibinfo {author} {\bibfnamefont {J.~J.}\ \bibnamefont
  {Sakurai}},\ }\href@noop {} {\emph {\bibinfo {title} {Modern Quantum
  Mechanics}}}\ (\bibinfo  {publisher} {Addison Wesley},\ \bibinfo {year}
  {1994})\BibitemShut {NoStop}%
\bibitem [{\citenamefont {Gardiner}\ and\ \citenamefont
  {Zoller}(2004)}]{Gardiner2004}%
  \BibitemOpen
  \bibfield  {author} {\bibinfo {author} {\bibfnamefont {C.}~\bibnamefont
  {Gardiner}}\ and\ \bibinfo {author} {\bibfnamefont {P.}~\bibnamefont
  {Zoller}},\ }\href@noop {} {\emph {\bibinfo {title} {Quantum noise: a
  handbook of Markovian and non-Markovian quantum stochastic methods with
  applications to quantum optics}}},\ Vol.~\bibinfo {volume} {56}\ (\bibinfo
  {publisher} {Springer},\ \bibinfo {year} {2004})\BibitemShut {NoStop}%
\bibitem [{\citenamefont {Spohn}(1978)}]{Spohn1978a}%
  \BibitemOpen
  \bibfield  {author} {\bibinfo {author} {\bibfnamefont {H.}~\bibnamefont
  {Spohn}},\ }\href {\doibase 10.1063/1.523789} {\bibfield  {journal} {\bibinfo
   {journal} {Journal of Mathematical Physics}\ }\textbf {\bibinfo {volume}
  {19}},\ \bibinfo {pages} {1227} (\bibinfo {year} {1978})}\BibitemShut
  {NoStop}%
\bibitem [{\citenamefont {Breuer}\ \emph {et~al.}(2000)\citenamefont {Breuer},
  \citenamefont {Huber},\ and\ \citenamefont {Petruccione}}]{Breuer2000}%
  \BibitemOpen
  \bibfield  {author} {\bibinfo {author} {\bibfnamefont {H.~P.}\ \bibnamefont
  {Breuer}}, \bibinfo {author} {\bibfnamefont {W.}~\bibnamefont {Huber}}, \
  and\ \bibinfo {author} {\bibfnamefont {F.}~\bibnamefont {Petruccione}},\
  }\href@noop {} {\bibfield  {journal} {\bibinfo  {journal} {Phys. Rev. E}\
  }\textbf {\bibinfo {volume} {61}},\ \bibinfo {pages} {4883} (\bibinfo {year}
  {2000})}\BibitemShut {NoStop}%
\bibitem [{\citenamefont {Kohn}(2001)}]{Kohn2001}%
  \BibitemOpen
  \bibfield  {author} {\bibinfo {author} {\bibfnamefont {W.}~\bibnamefont
  {Kohn}},\ }\href@noop {} {\bibfield  {journal} {\bibinfo  {journal} {J. Stat.
  Phy}\ }\textbf {\bibinfo {volume} {103}},\ \bibinfo {pages} {417} (\bibinfo
  {year} {2001})}\BibitemShut {NoStop}%
\bibitem [{\citenamefont {Langemeyer}\ and\ \citenamefont
  {Holthaus}(2014)}]{Langemeyer2014}%
  \BibitemOpen
  \bibfield  {author} {\bibinfo {author} {\bibfnamefont {M.}~\bibnamefont
  {Langemeyer}}\ and\ \bibinfo {author} {\bibfnamefont {M.}~\bibnamefont
  {Holthaus}},\ }\href@noop {} {\bibfield  {journal} {\bibinfo  {journal}
  {Phys. Rev. E}\ }\textbf {\bibinfo {volume} {89}},\ \bibinfo {pages} {012101}
  (\bibinfo {year} {2014})}\BibitemShut {NoStop}%
\bibitem [{\citenamefont {Murch}\ \emph {et~al.}(2013)\citenamefont {Murch},
  \citenamefont {Weber}, \citenamefont {Macklin},\ and\ \citenamefont
  {Siddiqi}}]{Murch2013}%
  \BibitemOpen
  \bibfield  {author} {\bibinfo {author} {\bibfnamefont {K.}~\bibnamefont
  {Murch}}, \bibinfo {author} {\bibfnamefont {S.}~\bibnamefont {Weber}},
  \bibinfo {author} {\bibfnamefont {C.}~\bibnamefont {Macklin}}, \ and\
  \bibinfo {author} {\bibfnamefont {I.}~\bibnamefont {Siddiqi}},\ }\href@noop
  {} {\bibfield  {journal} {\bibinfo  {journal} {Nature}\ }\textbf {\bibinfo
  {volume} {502}},\ \bibinfo {pages} {211} (\bibinfo {year}
  {2013})}\BibitemShut {NoStop}%
\bibitem [{\citenamefont {Sun}\ \emph {et~al.}(2013)\citenamefont {Sun},
  \citenamefont {Petrenko}, \citenamefont {Leghtas}, \citenamefont {Vlastakis},
  \citenamefont {Kirchmair}, \citenamefont {Sliwa}, \citenamefont {Narla},
  \citenamefont {Hatridge}, \citenamefont {Shankar},\ and\ \citenamefont
  {Blumoff}}]{Sun2013}%
  \BibitemOpen
  \bibfield  {author} {\bibinfo {author} {\bibfnamefont {L.}~\bibnamefont
  {Sun}}, \bibinfo {author} {\bibfnamefont {A.}~\bibnamefont {Petrenko}},
  \bibinfo {author} {\bibfnamefont {Z.}~\bibnamefont {Leghtas}}, \bibinfo
  {author} {\bibfnamefont {B.}~\bibnamefont {Vlastakis}}, \bibinfo {author}
  {\bibfnamefont {G.}~\bibnamefont {Kirchmair}}, \bibinfo {author}
  {\bibfnamefont {K.~M.}\ \bibnamefont {Sliwa}}, \bibinfo {author}
  {\bibfnamefont {A.}~\bibnamefont {Narla}}, \bibinfo {author} {\bibfnamefont
  {M.}~\bibnamefont {Hatridge}}, \bibinfo {author} {\bibfnamefont
  {S.}~\bibnamefont {Shankar}}, \ and\ \bibinfo {author} {\bibfnamefont
  {J.}~\bibnamefont {Blumoff}},\ }\href@noop {} {\bibfield  {journal} {\bibinfo
   {journal} {Nature}\ }\textbf {\bibinfo {volume} {511}},\ \bibinfo {pages}
  {444} (\bibinfo {year} {2013})}\BibitemShut {NoStop}%
\bibitem [{\citenamefont {Vool}\ \emph {et~al.}(2014)\citenamefont {Vool},
  \citenamefont {Pop}, \citenamefont {Sliwa}, \citenamefont {Abdo},
  \citenamefont {Wang}, \citenamefont {Brecht}, \citenamefont {Gao},
  \citenamefont {Shankar}, \citenamefont {Hatridge}, \citenamefont {Catelani},
  \citenamefont {Mirrahimi},\ and\ \citenamefont {Frunzio}}]{Vool2014}%
  \BibitemOpen
  \bibfield  {author} {\bibinfo {author} {\bibfnamefont {U.}~\bibnamefont
  {Vool}}, \bibinfo {author} {\bibfnamefont {I.~M.}\ \bibnamefont {Pop}},
  \bibinfo {author} {\bibfnamefont {K.}~\bibnamefont {Sliwa}}, \bibinfo
  {author} {\bibfnamefont {B.}~\bibnamefont {Abdo}}, \bibinfo {author}
  {\bibfnamefont {C.}~\bibnamefont {Wang}}, \bibinfo {author} {\bibfnamefont
  {T.}~\bibnamefont {Brecht}}, \bibinfo {author} {\bibfnamefont {Y.~Y.}\
  \bibnamefont {Gao}}, \bibinfo {author} {\bibfnamefont {S.}~\bibnamefont
  {Shankar}}, \bibinfo {author} {\bibfnamefont {M.}~\bibnamefont {Hatridge}},
  \bibinfo {author} {\bibfnamefont {G.}~\bibnamefont {Catelani}}, \bibinfo
  {author} {\bibfnamefont {M.}~\bibnamefont {Mirrahimi}}, \bibinfo
  {author} {\bibfnamefont {L.}\ \bibnamefont {Frunzio}}, \bibinfo{author}{
  \bibfnamefont {R.~L.}~\bibnamefont{Schoelkopf}},\ }\href@noop {} {\bibfield  {journal}
  {\bibinfo  {journal} {Phys. Rev. Lett.}\ }\textbf {\bibinfo {volume} {113}},\
  \bibinfo {pages} {247001} (\bibinfo {year} {2014})}\BibitemShut {NoStop}%
\bibitem [{\citenamefont {Campagne-Ibarcq}\ \emph {et~al.}(2016)\citenamefont
  {Campagne-Ibarcq}, \citenamefont {Six}, \citenamefont {Bretheau},
  \citenamefont {Sarlette}, \citenamefont {Mirrahimi}, \citenamefont
  {Rouchon},\ and\ \citenamefont {Huard}}]{Campagne-Ibarcq2016}%
  \BibitemOpen
  \bibfield  {author} {\bibinfo {author} {\bibfnamefont {P.}~\bibnamefont
  {Campagne-Ibarcq}}, \bibinfo {author} {\bibfnamefont {P.}~\bibnamefont
  {Six}}, \bibinfo {author} {\bibfnamefont {L.}~\bibnamefont {Bretheau}},
  \bibinfo {author} {\bibfnamefont {A.}~\bibnamefont {Sarlette}}, \bibinfo
  {author} {\bibfnamefont {M.}~\bibnamefont {Mirrahimi}}, \bibinfo {author}
  {\bibfnamefont {P.}~\bibnamefont {Rouchon}}, \ and\ \bibinfo {author}
  {\bibfnamefont {B.}~\bibnamefont {Huard}},\ }\href {\doibase
  10.1103/PhysRevX.6.011002} {\bibfield  {journal} {\bibinfo  {journal} {Phys.
  Rev. X}\ }\textbf {\bibinfo {volume} {6}},\ \bibinfo {pages} {011002}
  (\bibinfo {year} {2016})}\BibitemShut {NoStop}%
\bibitem [{\citenamefont {Breuer}(2004)}]{Breuer2004}%
  \BibitemOpen
  \bibfield  {author} {\bibinfo {author} {\bibfnamefont {H.-P.}\ \bibnamefont
  {Breuer}},\ }\href {\doibase 10.1103/PhysRevA.70.012106} {\bibfield
  {journal} {\bibinfo  {journal} {Phys. Rev. A}\ }\textbf {\bibinfo {volume}
  {70}},\ \bibinfo {pages} {012106} (\bibinfo {year} {2004})}\BibitemShut
  {NoStop}%
\bibitem [{\citenamefont {Hush}\ \emph {et~al.}(2015)\citenamefont {Hush},
  \citenamefont {Lesanovsky},\ and\ \citenamefont {Garrahan}}]{Hush2015}%
  \BibitemOpen
  \bibfield  {author} {\bibinfo {author} {\bibfnamefont {M.~R.}\ \bibnamefont
  {Hush}}, \bibinfo {author} {\bibfnamefont {I.}~\bibnamefont {Lesanovsky}}, \
  and\ \bibinfo {author} {\bibfnamefont {J.~P.}\ \bibnamefont {Garrahan}},\
  }\href {http://dx.doi.org/10.1103/PhysRevA.91.032113} {\bibfield  {journal}
  {\bibinfo  {journal} {Phy. Rev. A}\ }\textbf {\bibinfo {volume} {91}},\
  \bibinfo {pages} {032113} (\bibinfo {year} {2015})}\BibitemShut {NoStop}%
\bibitem [{\citenamefont {Piilo}\ \emph {et~al.}(2008)\citenamefont {Piilo},
  \citenamefont {Maniscalco}, \citenamefont {H\"ark\"onen},\ and\ \citenamefont
  {Suominen}}]{Piilo2008}%
  \BibitemOpen
  \bibfield  {author} {\bibinfo {author} {\bibfnamefont {J.}~\bibnamefont
  {Piilo}}, \bibinfo {author} {\bibfnamefont {S.}~\bibnamefont {Maniscalco}},
  \bibinfo {author} {\bibfnamefont {K.}~\bibnamefont {H\"ark\"onen}}, \ and\
  \bibinfo {author} {\bibfnamefont {K.-A.}\ \bibnamefont {Suominen}},\ }\href
  {\doibase 10.1103/PhysRevLett.100.180402} {\bibfield  {journal} {\bibinfo
  {journal} {Phys. Rev. Lett.}\ }\textbf {\bibinfo {volume} {100}},\ \bibinfo
  {pages} {180402} (\bibinfo {year} {2008})}\BibitemShut {NoStop}%
\bibitem [{\citenamefont {Daley}(2014)}]{Daley2014}%
  \BibitemOpen
  \bibfield  {author} {\bibinfo {author} {\bibfnamefont {A.~J.}\ \bibnamefont
  {Daley}},\ }\href@noop {} {\bibfield  {journal} {\bibinfo  {journal} {Adv.
  Phys}\ }\textbf {\bibinfo {volume} {63}},\ \bibinfo {pages} {77} (\bibinfo
  {year} {2014})}\BibitemShut {NoStop}%
\bibitem [{\citenamefont {Kac}(1949)}]{Kac1949}%
  \BibitemOpen
  \bibfield  {author} {\bibinfo {author} {\bibfnamefont {M.}~\bibnamefont
  {Kac}},\ }\href {\doibase 10.1090/s0002-9947-1949-0027960-x} {\bibfield
  {journal} {\bibinfo  {journal} {Trans. Am. Math. Soc.}\ }\textbf {\bibinfo
  {volume} {65}},\ \bibinfo {pages} {1} (\bibinfo {year} {1949})}\BibitemShut
  {NoStop}%
\bibitem [{\citenamefont {Liu}(2012{\natexlab{b}})}]{Liu2012a}%
  \BibitemOpen
  \bibfield  {author} {\bibinfo {author} {\bibfnamefont {F.}~\bibnamefont
  {Liu}},\ }\href {http://arxiv.org/abs/1210.5798} {\bibfield  {journal}
  {\bibinfo  {journal} {arXiv:1210.5798}\ } (\bibinfo {year}
  {2012}{\natexlab{b}})}\BibitemShut {NoStop}%
\end{thebibliography}

\end{document}